\documentclass[11pt,reqno]{amsart}
\usepackage{graphics,amsmath,amssymb,epsfig,stmaryrd,color,amsaddr}
\usepackage{subfigure,amsfonts,geometry,array}
\usepackage{graphicx}
\usepackage{listings}
\usepackage{upquote}
\usepackage{wrapfig}
\usepackage{amsthm}
\usepackage{amsmath}
\usepackage{flafter}
\usepackage{bbm}
\usepackage{lscape}
\usepackage{pdflscape}
\usepackage{booktabs, topcapt}
\usepackage{threeparttable}
\usepackage{lscape}
\usepackage{pdflscape}
\usepackage{mathrsfs}
\usepackage{hyperref}
\usepackage{cleveref}
\usepackage{pgfplots} 
\usepackage{natbib}
\usepackage{algorithm,algpseudocode}
\usepackage{enumitem}
\usepackage{enumerate}
\newcommand{\indep}{\rotatebox[origin=c]{90}{$\models$}}
\newtheorem{theorem}{Theorem}
\theoremstyle{plain}

\newtheorem{assumption}{Assumption}

\newtheorem{proposition}{Proposition}
\newtheorem{remark}{Remark}

\numberwithin{equation}{section}

\begin{document}
\nonstopmode

\title[Testing identifying assumptions in Tobit Models]{Testing identifying assumptions in Tobit models}
\author{Santiago Acerenza$^1$ Ot\'avio Bartalotti$^2$  and Federico Veneri$^3$
}
\address{$^1$Universidad ORT Uruguay $^2$Monash University, IZA $^3$Iowa State University} 
\noindent \thanks{\scriptsize{The present version is as of \today, the first version is from December 2022. All errors are ours. Email address: otavio.bartalotti@monash.edu}. We want to thank participants at seminars in the 2024 Jornadas Anuales de Economía del Banco Central del Uruguay and RMIT University, the 2024 IAAE and Southern Economic Association meetings.}

\begin{abstract}
We develop sharp, testable implications for the identifying assumptions of Tobit and IV-Tobit models: linear index, (joint) normality of errors, treatment (instrument) exogeneity, and relevance. The new sharp testable equalities can detect all possible observable violations of the identifying conditions. The proposed test procedure for the model's validity uses existing inference methods for intersection bounds. Simulations suggest adequate test size and power in detecting exogeneity and error structure violations. We review and propose alternatives to partially identify the parameters of interest under less restrictive assumptions. We revisit a study of married women's labor supply in \cite{lee1995semi} to demonstrate the test’s practical implementation.
\end{abstract}
\maketitle
{\footnotesize \textbf{Keywords}: Tobit models, hypothesis testing, Testable Implications, Instrumental variables.

\textbf{JEL subject classification}: C12, C24, C26, C34.}

\section{Introduction}\label{S1}
Since the seminal work of \cite{tobin_estimation_1958}, Tobit models have earned attention in economics, business and social sciences.\footnote{According to Google Scholar Tobin's original paper has more than 10000 citations.} \cite{tobin_estimation_1958} analyzed household expenditure on durable goods using a regression model that specifically incorporated that expenditure (the dependent variable) cannot be negative. This approach is related to a broader class of censored or truncated regression models, depending on whether observations outside a specified range are lost or censored. When applied researchers are interested in modelling limited dependent variables, potentially with mass accumulation points, the Tobit family of models provides structure to identify parameters of interest, such as the average treatment effect (ATE). Identification relies on three primary sources: (i) instrument exogeneity (or exogeneity of the variable of interest itself), (ii) normality of the model's latent variables, and, in the case of an instrumental variable approach to endogeneity, (iii) the relevance condition for the instrument. While researchers recognize the model's restrictive nature, it remains a valuable tool in the empirical literature.
\par 
 In this paper, we develop a test for the validity of the Tobit model's structure and assumptions, providing three main contributions to the literature. The first is to provide the set of sharp testable equalities that can detect all possible observable violations of the Tobit model. Second, we propose a test for the validity of the Tobit model's identifying assumptions using the sharp equalities that characterize the model to check its falsifiability. Following recent literature, we convert the equalities into conditional moment inequalities and implement the test by existing inferential methods from \cite{chernozhukov_intersection_2013}.
The Tobit model family is used for continuous outcomes with accumulation points. In the case of household expenditure, the outcome may exhibit a zero accumulation due to censoring. When evaluating causality, researchers may be interested in a treatment variable, not necessarily binary, which would result in a large number of moment equalities to be tested. This creates significant challenges to test implementation. We propose a discretization of the space of the treatment and the outcome that balances the computational requirements and data availability for different parts of their joint support. This simplifies the implementation, making it easy to compute and providing an asymptotically valid testing procedure.

The third contribution is to review and propose alternative approaches that can be used when the model is rejected. We explore an alternative path to partially identify the parameter of interest by assuming the monotonicity of selection into treatment. Finally,  we provide an empirical example illustrating the methodology's practical relevance. More generally, the current paper contributes to the growing literature on testing identifying assumptions of econometric models.
\par 
 We focus on two main models: (i) the ``classic Tobit'' model in which the main variable of interest is assumed to be exogenous and (ii) the instrumental variable (IV) Tobit model. In both cases, the proposed test considers all observable violations of the model structure, such as linear index, normality of the latent errors, independence of the treatment, and homoskedasticity; with the addition of the validity of the instrument for the IV-Tobit. Additional results for variants of the Tobit family of models are presented in the Appendix.

\subsection{Previous Literature} 
There is a vast literature related to testing the validity of Tobit models and their assumptions, which we contribute to. Most of the preceding work focuses on testing a particular assumption or feature of the model while maintaining other assumptions and structures as valid. \cite{nelson_test_1981} constructs a Hausman-type test for misspecification of the classic Tobit model (that is, normality, linear index, and homoskedasticity) where the maximum likelihood estimates are compared with method of moment estimates. Nelson's test compares the sample proportion of non-censored observations with the hypothesized probability of being non-censored in the Tobit model. \cite{bera_testing_1984} state that the test is equivalent to the Lagrange multiplier (LM) test of the Tobit model against Cragg's model \citep{lin_test_1984} and propose an alternative LM test for the normality assumption against other distributions of the Pearson family of distributions while the remaining assumptions are maintained. \cite{newey_specification_1987} considers both exogenous and endogenous explanatory variables cases using symmetrically censored least squares estimators to construct specification tests of normality and homoskedasticity assumptions via a Hausman-type specification test. \cite{holden_testing_2004} examines several statistics proposed to test the normality assumption in the Tobit (censored regression) model and reinterprets them as a version of the LM (score) test for a common null hypothesis. \cite{reynolds_testing_1991} use an information matrix misspecification test to detect violations of the distributional assumptions of the Tobit model.

Other tests include \cite{drukker_bootstrapping_2002}, which operationalized conditional moment tests developed by \cite{newey_maximum_1985} and \cite{tauchen_diagnostic_1985} to the case of misspecification of the distribution of the classic Tobit model. With a similar intuition to our framework, their test writes down conditional moment restrictions, which should have zero conditional expected values under the null. Since the model was estimated by maximum likelihood, the assumed data-generating process specifies the moments of disturbances conditional on the covariates to be the ones of a normal distribution. \cite{drukker_bootstrapping_2002} use these moment-based methods based on the third and fourth moments of the normal distribution. While intuitively similar, our procedure detects all possible violations of the model, not only those evident 
from deviations in the third and fourth moments. 
\cite{smith_exogeneity_1986} propose a test of the treatment variable's exogeneity in the IV-Tobit Model by a control function approach exploiting the joint normality of the latent variables. Most of these approaches focus on testing a particular subset of assumptions or consider a specific class of alternatives.

The developments proposed in this paper consider all observable violations of the general Tobit model structure and its assumptions, serving as a useful test for empirical researchers constructing models for censored/truncated data. This work contributes to the growing literature on the testability of the identifying assumptions in various econometric models. Those include sharp tests of the validity in instrumental variable models\footnote{\cite{gunsilius_nontestability_2021} proves that instrument validity cannot be tested in the case where the endogenous treatment is continuously distributed.} and local average effects \citep{mourifie_testing_2017,kitagawa_test_2015, huber_testing_2015,kedagni2020generalized}, sharp tests of the validity of assumption in the regression discontinuity designs \citep{arai_testing_2022}, sharp tests of the assumptions of the bivariate probit \citep{acerenza_testing_2023} and sharp tests in the context of encouragement designs \citep{bai2024sharptestableimplicationsencouragement} among others. Our testing procedures are connected to \cite{kedagni2020generalized}, which provides tests for assumptions related to the instrumental variables in the model, while we also consider the implications of the parametric structure on the outcome and latent error structure, which are inherent to Tobit models.

The two papers closest to ours are \cite{acerenza_testing_2023} and \cite{goff2024testing}. The first focuses on bivariate probit models, leading to a finite set of moment equalities as testable restrictions. As mentioned above, this study builds upon their work by tackling the case of the Tobit family of estimators, which provides additional technical and practical challenges given the continuous nature of outcome and treatment, including accumulation points.
\cite{goff2024testing} considers separable parametric instrumental variable models that induce a different set of sharp moment equalities to be tested while our approach focuses on Tobit and IV-Tobit models, which can potentially be non-separable.\footnote{While \cite{goff2024testing} briefly mentions the possibility of extending their results to non-separable models, they obtain different equalities from ours and do not discuss identification or sharpness in that context. Implementation would also differ from our approach.} 

The remainder of the paper is organized as follows. Section \ref{S2} presents the model and identifying assumptions.  Section \ref{S3} discusses heuristically the identification of the model's parameters. Section \ref{S9} derives the sharp testable implications. Section \ref{S4} outlines the testing procedure. Section \ref{S5} include simulation evidence about the test's size, while discussions about power are relegated to the Appendix. Section \ref{S6} discusses how to relax the assumptions in case of rejection. Section \ref{S7b} provides an empirical illustration revisiting the study of married women's labor supply from \cite{lee1995semi} to demonstrate the practical implementation and usefulness of the test. Finally, Section \ref{S8} concludes. Additional results and details are collected in Appendices \ref{app0}-\ref{ApLearn}.

\section{Models}\label{S2}
This section describes two popular ``two-part'' models for truncated data for which we derive testable implications in Section \ref{S9}: the classic Tobit, and the IV-Tobit.\footnote{We omit conditioning on other exogenous covariates ($X$) for simplicity of exposition. The results including covariates are presented in Section \ref{S43}.}

Let $Y=\max(0,Y^*)$ be the observed outcome taking values in $\mathcal{Y}\subset \mathbb{R}^{+}$ with $Y^*$ a latent continuous dependent variable taking values in $\mathcal{Y^*}\subset \mathbb{R}$. Both the treatment of interest, $D$, and the instrumental variable, $Z$, can be discrete or continuous and take values in $\mathcal{D} \subset \mathbb{R}$ and $\mathcal{Z}\subset \mathbb{R}$, respectively. We consider the normalized coefficients by the standard deviation of the unobservable error term, which is convenient for the exposition.\footnote{Identification for the normalized and non-normalized parameters is discussed in section \ref{S3}.} 
\subsection{Classic Tobit model}\label{S21}
The classic Tobit model considers the case in which the researcher is interested in the effect of an exogenous treatment on a non-negative outcome that has a mass point at zero:
 \begin{eqnarray}\label{CT1}
\left\{ \begin{array}{lcl}
     Y &=& \max(0,Y^*) \\ \\
     Y^* &=& \tilde{\alpha}_0+\tilde{\alpha}_1 D+U
     \end{array} \right.
\end{eqnarray}
where $U$ is an unobservable (latent) error. In addition to the model structure in system \ref{CT1}, the classic Tobit restricts the distribution of the error term and its relationship to the treatment.
\begin{assumption}\label{CT2}
$D$ is independent of $U$.
\end{assumption}
\begin{assumption}\label{CT3}
$U$ is distributed as $N(0,\sigma^2)$.
\end{assumption}
Assumption \ref{CT2} states that the treatment of interest is independent of the model's unobservables. Assumption \ref{CT3} imposes that the latent error has a normal distribution. To simplify some of the derivations later in the manuscript, we defined the scaled version the model for which the transformed error term would have distribution $N(0,1)$.
\begin{equation*}
     \frac{Y^*}{\sigma}=\frac{\tilde{\alpha}_0}{\sigma}+\frac{\tilde{\alpha}_1}{{\sigma}} D+\frac{U}{\sigma}=\alpha_0+\alpha_1 D+\frac{U}{\sigma}
\end{equation*}
We refer to the standardized parameters without the tilde, thus for example $\alpha_1=\frac{\tilde{\alpha}_1}{\sigma}$.

Despite its simplicity, the classic Tobit model can accommodate continuous or discrete treatments, and the normality assumption for the latent error can be replaced with other distributional assumptions that preserve parameter identification.

\subsection{IV Tobit model}\label{S22}
The IV Tobit considers the case in which the treatment of interest is endogenous, but an instrumental variable is available to identify its effect. It can be traced back to \cite{heckman1977dummy,amemiya1979estimation,nelson1978specification} among several approaches to consider censored variables with endogeneity.\footnote{In Section \ref{S6}, we discuss alternative models and how identification can still be achieved under different assumptions, including about the distribution of the latent error terms.} Let the IV Tobit model be:\footnote{Results for the IV-Tobit model are presented for continuous $D$ but hold for a binary treatment, $D=1\{\gamma_0+\gamma_1 Z-V>0\}$. In that case, the parameters are identified only up to scale.}
 \begin{eqnarray}\label{IVT0}
\left\{ \begin{array}{lcl}
     Y &=& \max(0,Y^*) \\
     Y^* &=& \tilde{\alpha}_0+\tilde{\alpha}_1 D+U  \\
     D&=& \tilde{\gamma}_0+\tilde{\gamma}_1 Z+\tilde{V}
     \end{array} \right.
\end{eqnarray}
Where $U, \tilde{V}$ are the latent structural error terms.
Alternatively, in its reduced form representation: 
\begin{eqnarray}\label{IVT1}
\left\{ \begin{array}{lcl}
     Y &=& \max(0,Y^*) \\
     Y^* &=& \tilde{\beta}_0+\tilde{\beta}_1 Z+\tilde{W}  \\
     D&=& \tilde{\gamma}_0+\tilde{\gamma}_1 Z+\tilde{V}
     \end{array} \right.
\end{eqnarray}
Where $\tilde{\beta}_0=\tilde{\alpha}_0+\tilde{\alpha}_1\tilde{\gamma}_0, \tilde{\beta}_1=\tilde{\alpha}_1 \tilde{\gamma}_1$ and  $\tilde{W}, \tilde{V}$ are the reduce form error terms where $\tilde{W}=\tilde{\alpha}_1\tilde{V}+U$. 
\par 
 To that model structure, restrictions on the joint distribution of the latent variables and their relationship to the instrumental variable are added.
\begin{assumption}\label{IVT2} 
$Z$ is independent of $(U,\tilde{V})$ and $\tilde{\gamma}_1 \neq 0$.

\end{assumption}
\begin{assumption}\label{IVT3}
Let $U, \tilde{V}$ follow a bivariate normal  distribution with covariance $\rho_{U\tilde{V}}$, i.e., $\begin{pmatrix}
U\\\tilde{V}
\end{pmatrix} \sim \mathcal N(\mu,\Sigma)$, where
$\mu=\begin{pmatrix}
0\\0\
\end{pmatrix}$, and 
$\Sigma=
\begin{pmatrix}
\sigma^2_U & \rho_{U\tilde{V}}  \\ 
 \rho_{U\tilde{V}} & \sigma^2_{\tilde{V}}
\end{pmatrix}$.\end{assumption}
Assumption \ref{IVT2}(i) states that the instrumental variable is independent of the model's structural latent variables. The second part of Assumption \ref{IVT2} is the usual instrument relevance in determining the treatment. This assumption also implies that $Z$ is independent of $\tilde{W}, \tilde{V}$, since $\tilde{W}=\tilde{\alpha}_1\tilde{V}+U$.
\par 
Assumption \ref{IVT3} characterizes the distribution of the latent vector of error terms as bivariate normal. Then, $\tilde{W}, \tilde{V}$ also follows a bivariate normal distribution and $\tilde{\alpha}_0, \tilde{\alpha}_1, \tilde{\gamma}_0, \tilde{\gamma}_1$ could be scale normalized so  $(W,V)'$ a normalization of $(\tilde{W},\tilde{V})'$  follows a standard bivariate normal distribution with covariance $\rho$, i.e., $\begin{pmatrix}
W\\V
\end{pmatrix} \sim \mathcal N(\mu,\Sigma)$, where
$\mu=\begin{pmatrix}
0\\0
\end{pmatrix}$, and 
$\Sigma=
\begin{pmatrix}
1 & \rho  \\ 
 \rho & 1
\end{pmatrix}$.\footnote{{Note that if  $U, \tilde{V}$ follow a bivariate normal distribution, $\begin{pmatrix}
U\\\tilde{V}
\end{pmatrix} \sim \mathcal N(\mu,\Sigma)$, where
$\mu=\begin{pmatrix}
0\\0\
\end{pmatrix}$, and 
$\Sigma=
\begin{pmatrix}
\sigma^2_U & \rho_{U\tilde{V}}  \\ 
 \rho_{U\tilde{V}} & \sigma^2_{\tilde{V}}
\end{pmatrix}$, then  $\tilde{W}, \tilde{V}$ follow a bivariate normal  distribution, $\begin{pmatrix}
\tilde{W}\\\tilde{V}
\end{pmatrix} \sim \mathcal N(\mu_1,\Sigma_1)$, where
$\mu_1=\begin{pmatrix}
0\\0\
\end{pmatrix}$, and 
$\Sigma_1=
\begin{pmatrix}
\tilde{\alpha}_1^2\sigma^2_{\tilde{V}}+\sigma^2_U+2\rho_{U\tilde{V}}\tilde{\alpha}_1\sigma_{\tilde{V}}\sigma_U  & \tilde{\alpha}_1 \sigma_{\tilde{V}}^2+\rho_{U\tilde{V}}  \\ 
\tilde{\alpha}_1 \sigma_{\tilde{V}}^2+\rho_{U\tilde{V}} & \sigma^2_{\tilde{V}}
\end{pmatrix}\equiv \begin{pmatrix}
\sigma^2_{\tilde{W}}  & \rho_{\tilde{W}\tilde{V}}   \\ 
 \rho_{\tilde{W}\tilde{V}} &  \sigma^2_{\tilde{V}}
\end{pmatrix} $. Then the reduced form system can be scale normalized to   
$W=\frac{\tilde{W}}{\sigma_{\tilde{W}}}, V=\frac{\tilde{V}}{\sigma_{\tilde{V}}}$ which follows a bivariate normal  distribution, $\begin{pmatrix}
W\\V
\end{pmatrix} \sim \mathcal N(\mu_2,\Sigma_2)$, where
$\mu_2=\begin{pmatrix}
0\\0\
\end{pmatrix}$, and 
$\Sigma_2=
\begin{pmatrix}
1  & \frac{\rho_{\tilde{W}\tilde{V}}}{\sigma_{\tilde{W}}\sigma_{\tilde{V}}}   \\ 
 \frac{\rho_{\tilde{W}\tilde{V}}}{\sigma_{\tilde{W}}\sigma_{\tilde{V}}} &  1
\end{pmatrix}\equiv  \begin{pmatrix}
1  & \rho   \\ 
 \rho &  1
\end{pmatrix}$.Define analogously, $\beta_0=\frac{\tilde{\beta}_0}{\sigma_{\tilde{W}}}, \beta_1=\frac{\tilde{\beta}_1}{\sigma_{\tilde{W}}}, \gamma_0=\frac{\tilde{\gamma}_0}{\sigma_{\tilde{V}}}, \gamma_1=\frac{\tilde{\gamma}_1}{\sigma_{\tilde{V}}}$.}
}  The latter characterization aids the development of the sharp testable implications. 

\section{Identification} \label{S3} 
In this section, we provide heuristic arguments for the identification of the parameters of the respective models, expanding on \cite{han_identification_2017} and \cite{acerenza_testing_2023}. These are known in the literature and discussed here for completeness and intuition. The testable implications that are our main result are presented in Section \ref{S9}.
\subsection{Classic Tobit model}\label{S31}
Note that, 
\begin{eqnarray*}
1-P(Y=0|D)=1-P(Y^{*}\leq 0|D)=\Phi(\alpha_0+\alpha_1 D).
\end{eqnarray*}
The equality follows from the model's structure in Equation \ref{CT1} and assumptions \ref{CT2} and \ref{CT3}. 

Since the standard normal CDF, $\Phi(.)$, is monotonic, we use its inverse to obtain: 
\begin{eqnarray*}
\Phi^{-1}(1-P(Y=0|D))=\alpha_0+\alpha_1 D.
\end{eqnarray*}
Thus, $\alpha_1=\frac{Cov(\Phi^{-1}(1-P(Y=0|D)),D)}{Var(D)}$, and $\alpha_0=E(\Phi^{-1}(1-P(Y=0|D)))-\alpha_1 E(D)$.
\par 
The coefficients have been normalized by dividing by the latent variable's ($U$) square root of the variance, denoted by $\sigma$.  This implies that the ratio of the original parameters is identified since $\frac{\alpha_1}{\alpha_0}=\frac{\frac{\tilde{\alpha}_1}{\sigma}}{\frac{\tilde{\alpha}_0}{\sigma}}= \frac{\tilde{\alpha}_1}{\tilde{\alpha}_0}$.  
 {
Recall that,
\begin{eqnarray*}
F_{U|U<c}(u|U<c)=P(U \leq u|U<c)&=& \frac{P(U \leq \min(u,c))}{P(U< c)}=\begin{cases}
 \frac{P(U \leq c)}{P(U\leq c)}=1, \text{ for } u \geq c\\
 \frac{P(U \leq u)}{P(U\leq c)}=\frac{\Phi(u)}{\Phi(c)}, \text{ for } u < c    
\end{cases}  
\end{eqnarray*}
Then, $f_{U|U<c}(u|U<c)=\begin{cases}
 0, \text{ for } u \geq c\\
 \frac{\phi(u)}{\Phi(c)}, \text{ for } u < c    
\end{cases}  
$, and $f_{U|U>c}(u|U>c)=\begin{cases}
 0,\text{ for } u \leq c \\
 \frac{\phi(u)}{1-\Phi(c)},\text{ for } u > c     
\end{cases}  
$.
Note then that, 
\begin{eqnarray*}
E(Y|D,Y>0)&=& E(\tilde{\alpha}_0+\tilde{\alpha}_1 D+U|D,Y>0) \\
&=& \tilde{\alpha}_0+\tilde{\alpha}_1 D+E(U|D,Y>0) \\
&=&  \tilde{\alpha}_0+\tilde{\alpha}_1 D+E(U|D,\tilde{\alpha}_0+\tilde{\alpha}_1 D+U>0) \\
&=& \tilde{\alpha}_0+\tilde{\alpha}_1 D+E(U|U>-\tilde{\alpha}_0-\tilde{\alpha}_1 D) \\
&=& \tilde{\alpha}_0+\tilde{\alpha}_1 D+\sigma \frac{\phi(- \frac{\tilde{\alpha}_0}{\sigma}- \frac{\tilde{\alpha}_1}{\sigma} D)}{1-\Phi(- \frac{\tilde{\alpha}_0}{\sigma}- \frac{\tilde{\alpha}_1}{\sigma} D)}
\end{eqnarray*}
where the fourth equality uses independence, and the fifth equality uses the properties of standard truncated normal distributions.
Since $\alpha_0,\alpha_1$ are identified, then the inverse Mills ratio is identified,
\begin{eqnarray*}
 \frac{\phi(- \frac{\tilde{\alpha}_0}{\sigma}- \frac{\tilde{\alpha}_1}{\sigma} D)}{1-\Phi(- \frac{\tilde{\alpha}_0}{\sigma}- \frac{\tilde{\alpha}_1}{\sigma} D)}&=& \lambda(\tilde{\alpha}_0/\sigma+\tilde{\alpha}_1/\sigma D)= \lambda(\alpha_0+\alpha_1 D)    
\end{eqnarray*}
\par 
Thus, 
}
\begin{eqnarray*}
E(Y|D,Y>0)=\tilde{\alpha}_0+\tilde{\alpha}_1 D+\sigma \lambda(\alpha_0+\alpha_1 D)
\end{eqnarray*} 
Which in turn implies (by the property that we can express $X=E(X|W)+e$, where  $E(e|W)=0$): 
\begin{eqnarray*}
Y=\tilde{\alpha}_0+\tilde{\alpha}_1 D+\sigma\lambda(\alpha_0+\alpha_{1} D)+e
\end{eqnarray*} 
with $e$ mean independent of $D, \lambda(\alpha_0+\alpha_1 D)$. Since $D$ is known and $\lambda(\cdot)$ is identified, we get a three-by-three linear system with a unique solution for $\tilde{\alpha}_0, \tilde{\alpha}_1, \sigma$. In other words, the normality assumption allows identification by leveraging the relationship of the variable of interest ($D$) and the outcome at the accumulation point ($Y=0$) to obtain the scaled parameters, and the relationship of treatment with the distribution of $Y$ (when $Y>0$) to overcome the normalization. However, the reliance on the normality and linearity assumptions underlines the importance of adequately testing the model's structure and assumptions in empirical applications.

\subsection{IV Tobit model}\label{S32}
Turning the focus to the IV-Tobit model, the first stage is readily identified from the linear structure, $\tilde{\gamma}_1=\frac{Cov(Z,D)}{Var(Z)}$, $\tilde{\gamma}_0=E(D)-\tilde{\gamma}_1 E(Z)$. 
{ Furthermore, $\tilde{V}=D-E(D)+\frac{Cov(Z,D)}{Var(Z)} E(Z)-\frac{Cov(Z,D)}{Var(Z)}Z$, thus $\tilde{V}$ and $\sigma_{\tilde{V}}^2$ are identified. Combining these results implies that we can identify $\gamma_0=\frac{\tilde{\gamma}_0}{\sigma_{\tilde{V}}}$ and $\gamma_1=\frac{\tilde{\gamma}_1}{\sigma_{\tilde{V}}}$.}
\par 
Now note that calculating $P(Y=0|Z)$ from the reduce form:
\begin{eqnarray*}
1-P(Y=0|Z)=\Phi(-\beta_0-\beta_1 Z)
\end{eqnarray*}
{
Where $\beta_0=\frac{\tilde{\beta}_0}{\sigma_{\tilde{W}}}$ and $\beta_1=\frac{\tilde{\beta}_1}{\sigma_{\tilde{W}}}$.}
\par 

Similarly to subsection \ref{S31}, by inverting the normal CDF we obtain $-\beta_1=\frac{Cov(\Phi^{-1}(1-P(Y=0|Z)),Z)}{Var(Z)}$, $-\beta_0=E(\Phi^{-1}(1-P(Y=0|Z)))+\beta_1 E(Z)$. { The relationship between $\beta_0, \beta_1, \gamma_0, \gamma_1, \sigma_{\tilde{V}}$ identifies $\tilde{\alpha}_0, \tilde{\alpha}_1$ up to scale $\left(\alpha_0=\frac{\tilde{\alpha}_0}{\sigma_{\tilde{W}}}, \alpha_1=\frac{\tilde{\alpha}_1}{\sigma_{\tilde{W}}}\right)$.} 
\par 
To identify $\rho$, let $s=c-\beta_0-\beta_1 z$ and $t=d-\gamma_0-\gamma_1 z$. Then, by using 
\begin{eqnarray*}
P(c_0\leq Y \leq c_1, d_0 \leq D \leq d_1| Z=z)&=& \Phi_{W,V}\left(\frac{c_1}{\sigma_{\tilde{W}}}-\beta_0-\beta_1 z,\frac{d_1}{\sigma_{\tilde{V}}}-\gamma_0-\gamma_1 z;\rho\right) \nonumber\\
&-&\Phi_{W,V}\left( \frac{c_0}{\sigma_{\tilde{W}}}-\beta_0-\beta_1 z,\frac{d_1}{\sigma_{\tilde{V}}}-\gamma_0-\gamma_1 z;\rho\right) \nonumber \\ 
&-& \Phi_{W,V}\left( \frac{c_1}{\sigma_{\tilde{W}}}-\beta_0-\beta_1 z,\frac{d_0}{\sigma_{\tilde{V}}}-\gamma_0-\gamma_1 z;\rho\right) \nonumber \\
&+&\Phi_{W,V}\left( \frac{c_0}{\sigma_{\tilde{W}}}-\beta_0-\beta_1 z,\frac{d_0}{\sigma_{\tilde{V}}}-\gamma_0-\gamma_1 z;\rho\right) \nonumber \\
&\equiv& \Phi^{1}_{c_{1},c_{0},d_{1},d_{0}}, 
\end{eqnarray*}
We can obtain 
\begin{eqnarray*}
P(Y \leq c, D \leq d| Z=z)&=& \Phi_{W,V}(s,t; \rho),
\end{eqnarray*}
which by classic results of bivariate normal random variables if we differentiate w.r.t. $\rho$ is: 
\begin{eqnarray*}
\frac{ \partial P(Y \leq c, D \leq d| Z=z)}{\partial \rho}&=& \phi_{W,V}(s,t; \rho)
\end{eqnarray*}
Where $\phi_{a,b}$ is the bivariate normal probability density, which is positive for any $s,t$. Hence, $\Phi_{W,V}(.,.; \rho)$ is monotonic in $\rho$ and thus invertible. So $\rho$ can be identified. 

Recall the previous coefficients are normalized, as noted by \cite[p.~683-84]{wooldridge_econometric_2010} we can identify the original and reduced form covariance matrices, as well as the true coefficients without the normalization.   
{
To explain this in detail, recall that we have identified $\rho$, the covariance between $W,V$ the parameters  $\gamma_1,\gamma_0,\beta_1,\beta_0$ and the variance of $\tilde{V}$, $\sigma^2_{\tilde{V}}$. Similarly to the classic tobit model, we can leverage normality of $W$, independence of $Z$, and exploit the reduced form representation of $Y$ to obtain:  
\begin{eqnarray*}
E(Y|Z,Y>0)&=&  \tilde{\beta}_0+\tilde{\beta}_1 Z+\sigma_{\tilde{W}} \lambda(\beta_0+\beta_1 Z)
\end{eqnarray*} 
Which in turn implies
\begin{eqnarray*}
Y=\tilde{\beta}_0+\tilde{\beta}_1 Z+\sigma_{\tilde{W}} \lambda(\beta_0+\beta_1 Z)+e
\end{eqnarray*} 
with $e$ mean independent of $(Z, \lambda(\beta_0+\beta_1 Z))$. Since $Z$ is known and $\lambda(\beta_0+\beta_1 z)$ is identified for any given $Z=z$, we can get a solution for $\sigma_{\tilde{W}}$ as one of the coefficients for the linear projection of $Y$ on $(Z, \lambda(\beta_0+\beta_1 Z))$, identifying the non-standardized reduced form error for $Y$. Finally, since $\beta_0=\frac{\tilde{\beta}_0}{\sigma_{\tilde{W}}}=\frac{\tilde{\alpha}_0+\tilde{\alpha}_1 \tilde{\gamma}_0}{\sigma_{\tilde{W}}}$, $\beta_1=\frac{\tilde{\beta}_1}{\sigma_{\tilde{W}}}=\frac{\tilde{\alpha}_1 \tilde{\gamma}_0}{\sigma_{\tilde{W}}}$ and we have identified $\sigma_{\tilde{V}}, \gamma_0, \gamma_1, \tilde{\gamma}_0, \tilde{\gamma}_1$ and $\sigma_{\tilde{W}}$ we can recover $\tilde{\alpha}_0, \tilde{\alpha}_1$. 
\par 
It now only remains to recover the original variance-covariance structure. Under bivariate normality of the latent variables in the structural model, we can express the following relationships between $U,\tilde{V}$ with $\tilde{W}=\tilde{\alpha}_1U+\tilde{V}, V$ and $W=\frac{\tilde{\alpha}_1U+\tilde{V}}{\sigma_{\tilde{W}}}, V=\frac{\tilde{V}}{\sigma_{\tilde{V}}}$\par
$\begin{pmatrix}
U\\\tilde{V}
\end{pmatrix} \sim \mathcal N\Bigg(\begin{pmatrix}
0\\0\
\end{pmatrix},\begin{pmatrix}
\sigma^2_U & \rho_{U\tilde{V}}  \\ 
 \rho_{U\tilde{V}} & \sigma^2_{\tilde{V}}
\end{pmatrix}\Bigg)$
\par 
$\begin{pmatrix}
\tilde{W}\\\tilde{V}
\end{pmatrix} \sim \mathcal N\Bigg(\begin{pmatrix}
0\\0\
\end{pmatrix},\begin{pmatrix}
\tilde{\alpha}_1^2\sigma^2_{\tilde{V}}+\sigma^2_U+2\rho_{U\tilde{V}}\tilde{\alpha}_1\sigma_{\tilde{V}}\sigma_U  & \tilde{\alpha}_1 \sigma_{\tilde{V}}^2+\rho_{U\tilde{V}}  \\ 
\tilde{\alpha}_1 \sigma_{\tilde{V}}^2+\rho_{U\tilde{V}} & \sigma^2_{\tilde{V}}
\end{pmatrix}\Bigg)$ 
\par 
 $\begin{pmatrix}
W\\V
\end{pmatrix} \sim \mathcal{N}\left(\begin{pmatrix}
0\\0\
\end{pmatrix},\begin{pmatrix}
1  & \frac{\tilde{\alpha}_1 \sigma_{\tilde{V}}^2+\rho_{U\tilde{V}} }{\big(\tilde{\alpha}_1^2\sigma^2_{\tilde{V}}+\sigma^2_U+2\rho_{U\tilde{V}}\tilde{\alpha}_1\sigma_{\tilde{V}}\sigma_U \big)\sigma_{\tilde{V}}}   \\ 
 \frac{\tilde{\alpha}_1 \sigma_{\tilde{V}}^2+\rho_{U\tilde{V}} }{\big(\tilde{\alpha}_1^2\sigma^2_{\tilde{V}}+\sigma^2_U+2\rho_{U\tilde{V}}\tilde{\alpha}_1\sigma_{\tilde{V}}\sigma_U \big)\sigma_{\tilde{V}}} &  1
\end{pmatrix}\right)$
\par 
Since we have identified $\tilde{\alpha}_1, \sigma_{\tilde{V}}, \sigma_{\tilde{W}}, \rho$  from the following equations: 
$$\rho=\frac{\tilde{\alpha}_1 \sigma_{\tilde{V}}^2+\rho_{U\tilde{V}} }{\big(\tilde{\alpha}_1^2\sigma^2_{\tilde{V}}+\sigma^2_U+2\rho_{U\tilde{V}}\tilde{\alpha}_1\sigma_{\tilde{V}}\sigma_U \big)\sigma_{\tilde{V}}}  $$
$$\sigma^2_{\tilde{W}}=\tilde{\alpha}_1^2\sigma^2_{\tilde{V}}+\sigma^2_U+2\rho_{U\tilde{V}}\tilde{\alpha}_1\sigma_{\tilde{V}}\sigma_U $$
We can identify $\sigma_{U}$ and $\rho_{U\tilde{V}}$, and thus, recover all the structural parameters.
}
\section{Sharp testable equalities} \label{S9} 
This section presents the sharp testable equalities for the models described in Section \ref{S2}. Deviations from these equalities imply violations of the ``null hypothesis'' that the Tobit model (linear latent index, treatment/instrument independence, instrument relevance, and normality) is valid. The equalities are conditional on parameters being identified under the model's assumptions, and a discussion on identification is postponed to Section \ref{S3}.  
\subsection{Classic Tobit model}\label{S91}
A defining characteristic of Tobit and similar models is the mass accumulation at zero for the distribution of the non-negative outcome, $Y$. Thus, the model's testable implications require characterizing the distribution at both the mass point and beyond it. 
\par 
Starting at the continuous part of the distribution of $Y$, the conditional probabilities of $c_0 \leq Y\leq c_1$ are observed for $c_1,c_0>0$. For any value of the treatment variables $d \in \mathcal{D}$:   
\begin{eqnarray}\label{CTEQ2}
P(c_0 \leq Y\leq c_1|D=d)&=&P(c_0 \leq Y\leq c_1,Y^* \geq 0|D=d)  \nonumber \\
&+&P(c_0 \leq Y\leq c_1,Y^*< 0|D=d) \nonumber \\
&=& P(c_0 \leq Y\leq c_1,Y^*\geq 0|D=d) \nonumber \\
&=& P(c_0 \leq Y^*\leq c_1,Y^* \geq 0|D=d)  \nonumber \\
&=& P(c_0 \leq Y^*\leq c_1|D=d) \nonumber \\
&=&P(c_0-\tilde{\alpha}_0-\tilde{\alpha}_1 d\leq U\leq c_1-\tilde{\alpha}_0-\tilde{\alpha}_1 d|D=d)  \nonumber \\
&=& \Phi\left(\frac{c_1}{\sigma}-\frac{\tilde{\alpha}_0}{\sigma}-\frac{\tilde{\alpha}_1}{\sigma} d\right)-\Phi\left(\frac{c_0}{\sigma}-\frac{\tilde{\alpha}_0}{\sigma}-\frac{\tilde{\alpha}_1}{\sigma} d\right)
\end{eqnarray}
where $\Phi(.)$ is the standard normal CDF. The first equality follows from the law of total probability. The second through fourth equalities follow from the model's structure described in Equation \ref{CT1}, namely $P(Y>0, Y^*< 0|D=d)=0$ and $Y^{*}=Y$ for $Y^{*}>0$. The fifth one is given by the latent linear model structure of $Y^*$, and finally, the last step follows from assumptions \ref{CT2} and \ref{CT3}.
\par 
Recalling the accumulation point, the observed event of $Y=0$ has a probability,
\begin{eqnarray}\label{CTEQ1}
P(Y=0|D=d)=P(Y^* \leq 0|D=d)=\Phi\left(-\frac{\tilde{\alpha}_0}{\sigma}-\frac{\tilde{\alpha}_1}{\sigma} d\right).
\end{eqnarray}
The equalities described in equations \ref{CTEQ2}-\ref{CTEQ1} fully characterize the distribution of $Y$ conditional on $D$, connecting the probabilities in the observed data to those implied by the Tobit model.  We collect these results more formally in Theorem \ref{Rem:SharpClassic}.
\begin{theorem}\label{Rem:SharpClassic}
Suppose that the classic Tobit model (\ref{CT1}) along with assumptions \ref{CT2} and \ref{CT3} hold. Then, the parameters $\tilde{\alpha}_{0}$, $\tilde{\alpha}_{1}$, $\sigma^{2}$ are identified, and equalities  (\ref{CTEQ2})-(\ref{CTEQ1}) must hold for all $c,d \in \mathcal{Y} \times \mathcal{D}$. Furthermore, these equalities are sharp, that is, whenever they hold, there exists a vector of $(\tilde{Y},D,\tilde{U})$ that satisfies model (\ref{CT1}), Assumptions \ref{CT2} and \ref{CT3}, and induces the observed distribution on the data $(Y,D)$.
\end{theorem}
The proof for the equalities follows from the discussion above, and further details about sharpness are presented in Appendix \ref{app0}.
Then, the equalities (\ref{CTEQ2})-(\ref{CTEQ1}) are sharp testable implications for the validity of the classic Tobit Model in \ref{CT1} coupled with assumptions \ref{CT2}-\ref{CT3}. They serve as the basis for the test procedure described in Section \ref{S4}.
\begin{proposition}[Non-learnability]\label{Rem:NLClassic}
The testable implications and sharpness discussed above show that the classic Tobit model is generally refutable. However, the model is non-verifiable in the sense that we can always construct a joint probability law of $(\Bar{Y}, D,\Bar{U})$ that violates the Tobit model validity but satisfies equalities (\ref{CTEQ2})-(\ref{CTEQ1}).  See Appendix \ref{ApLearn} for the proof.
\end{proposition}
\begin{proposition}[Extensions]\label{RemEx}
 The previous derivation can be adjusted for different variations of Tobit models, such as generalizations of the distributional assumptions \citep{barros2018generalized}, different thresholds \citep{carson2007Tobit}, dynamic Tobit models, or including individual-specific effects \citep{wooldridge2005simple, honore2000estimation, honore1993orthogonality}. In particular, note that similar approaches to the ones proposed in Theorem \ref{Rem:SharpClassic} can be used even if the latent errors don't follow a normal distribution and $Y^{*}$ does not have a linear index form, as long as the model is identified. In appendix \ref{app4}, we derive the equalities for the type 2 Tobit model, and similar logic can be applied to other two-part models. Additional testable results for the aforementioned models are discussed in Appendix \ref{ApRemEx}.
\end{proposition}
\subsection{IV Tobit model}\label{S92}
We turn our attention to the IV Tobit case and propose testable implications that can be used to test the model described in \ref{IVT0} and the associated assumptions \ref{IVT2} and \ref{IVT3}.
The observed data includes $(Y,D,Z)$, and we characterize the joint distribution of $(Y,D) \in \mathcal{Y} \times \mathcal{D}$ conditional on the instrument, $Z$, to obtain the model's testable implications. The mapping from observed probabilities to their corresponding quantities implied by the IV Tobit model requires jointly evaluating the continuous (interior) support for the outcome and treatment variables, as well as the accumulation points and distribution tails conditional on $Z$. Consider any $0<c_0<c_1,d_{0}<d_{1}$, and $z \in \mathcal{Y} \times\mathcal{D} \times \mathcal{Z}$,
\pagebreak
\begin{eqnarray}\label{IVTEQ1}
P(c_0\leq Y \leq c_1, d_0 \leq D \leq d_1| Z=z)&=& \Phi_{W,V}( \frac{c_1}{\sigma_{\tilde{W}}}-\frac{\tilde{\beta}_0}{\sigma_{\tilde{W}}}-\frac{\tilde{\beta}_1}{\sigma_{\tilde{W}}} z,\frac{d_1}{\sigma_{\tilde{V}}}-\frac{\tilde{\gamma_0}}{\sigma_{\tilde{V}}}-\frac{\tilde{\gamma_1}}{\sigma_{\tilde{V}}} z;\rho) \nonumber\\
&-&\Phi_{W,V}( \frac{c_0}{\sigma_{\tilde{W}}}-\frac{\tilde{\beta}_0}{\sigma_{\tilde{W}}}-\frac{\tilde{\beta}_1}{\sigma_{\tilde{W}}} z,\frac{d_1}{\sigma_{\tilde{V}}}-\frac{\tilde{\gamma_0}}{\sigma_{\tilde{V}}}-\frac{\tilde{\gamma_1}}{\sigma_{\tilde{V}}} z;\rho) \nonumber \\ 
&-& \Phi_{W,V}( \frac{c_1}{\sigma_{\tilde{W}}}-\frac{\tilde{\beta}_0}{\sigma_{\tilde{W}}}-\frac{\tilde{\beta}_1}{\sigma_{\tilde{W}}} z,\frac{d_0}{\sigma_{\tilde{V}}}-\frac{\tilde{\gamma_0}}{\sigma_{\tilde{V}}}-\frac{\tilde{\gamma_1}}{\sigma_{\tilde{V}}} z;\rho) \nonumber \\
&+&\Phi_{W,V}( \frac{c_0}{\sigma_{\tilde{W}}}-\frac{\tilde{\beta}_0}{\sigma_{\tilde{W}}}-\frac{\tilde{\beta}_1}{\sigma_{\tilde{W}}} z,\frac{d_0}{\sigma_{\tilde{V}}}-\frac{\tilde{\gamma_0}}{\sigma_{\tilde{V}}}-\frac{\tilde{\gamma_1}}{\sigma_{\tilde{V}}} z;\rho) \nonumber \\
&\equiv& \Phi^{1}_{c_{1},c_{0},d_{1},d_{0}}, \label{IVTEQ1a}
\end{eqnarray} 
\begin{eqnarray}
P(c_0\leq Y \leq c_1, D \leq d_0 | Z=z)
&=& \Phi_{W,V}\left( \frac{c_1}{\sigma_{\tilde{W}}}-\frac{\tilde{\beta}_0}{\sigma_{\tilde{W}}}-\frac{\tilde{\beta}_1}{\sigma_{\tilde{W}}} z,\frac{d_0}{\sigma_{\tilde{V}}}
-\frac{\tilde{\gamma_0}}{\sigma_{\tilde{V}}}-\frac{\tilde{\gamma_1}}{\sigma_{\tilde{V}}} z;\rho\right) \nonumber \\
&-&\Phi_{W,V}\left( \frac{c_0}{\sigma_{\tilde{W}}}-\frac{\tilde{\beta}_0}{\sigma_{\tilde{W}}}-\frac{\tilde{\beta}_1}{\sigma_{\tilde{W}}} z,\frac{d_0}{\sigma_{\tilde{V}}}
-\frac{\tilde{\gamma_0}}{\sigma_{\tilde{V}}}-\frac{\tilde{\gamma_1}}{\sigma_{\tilde{V}}} z;\rho\right) \nonumber \\
&\equiv& \Phi^{1}_{c_{1},c_{0},d_0}, \label{IVTEQ1b} \\
P(c_0\leq Y \leq c_1, D \geq d_1 | Z=z)&=& \Phi_{W}( \frac{c_1}{\sigma_{\tilde{W}}}-\frac{\tilde{\beta}_0}{\sigma_{\tilde{W}}}-\frac{\tilde{\beta}_1}{\sigma_{\tilde{W}}} z)-\Phi_{W}( \frac{c_0}{\sigma_{\tilde{W}}}-\frac{\tilde{\beta}_0}{\sigma_{\tilde{W}}}-\frac{\tilde{\beta}_1}{\sigma_{\tilde{W}}} z) \nonumber \\
&-& \Phi_{W,V}( \frac{c_1}{\sigma_{\tilde{W}}}-\frac{\tilde{\beta}_0}{\sigma_{\tilde{W}}}-\frac{\tilde{\beta}_1}{\sigma_{\tilde{W}}} z,\frac{d_1}{\sigma_{\tilde{V}}}-\frac{\tilde{\gamma_0}}{\sigma_{\tilde{V}}}-\frac{\tilde{\gamma_1}}{\sigma_{\tilde{V}}} z;\rho) \nonumber \\
&+& \Phi_{W,V}( \frac{c_0}{\sigma_{\tilde{W}}}-\frac{\tilde{\beta}_0}{\sigma_{\tilde{W}}}-\frac{\tilde{\beta}_1}{\sigma_{\tilde{W}}} z,\frac{d_1}{\sigma_{\tilde{V}}}-\frac{\tilde{\gamma_0}}{\sigma_{\tilde{V}}}-\frac{\tilde{\gamma_1}}{\sigma_{\tilde{V}}} z;\rho) \nonumber \\
&\equiv& \Phi^{1}_{c_{1},c_{0},d_{1}}, \label{IVTEQ1d}\\
P(c_1\leq Y, d_0 \leq D \leq d_1| Z=z) &=&  \Phi_{V}(\frac{d_1}{\sigma_{\tilde{V}}}-\frac{\tilde{\gamma_0}}{\sigma_{\tilde{V}}}-\frac{\tilde{\gamma_1}}{\sigma_{\tilde{V}}} z)-\Phi_{V}(\frac{d_0}{\sigma_{\tilde{V}}}
-\frac{\tilde{\gamma_0}}{\sigma_{\tilde{V}}}-\frac{\tilde{\gamma_1}}{\sigma_{\tilde{V}}} z) \nonumber \\
&-&\Phi_{W,V}( \frac{c_1}{\sigma_{\tilde{W}}}-\frac{\tilde{\beta}_0}{\sigma_{\tilde{W}}}-\frac{\tilde{\beta}_1}{\sigma_{\tilde{W}}} z,\frac{d_1}{\sigma_{\tilde{V}}}-\frac{\tilde{\gamma_0}}{\sigma_{\tilde{V}}}-\frac{\tilde{\gamma_1}}{\sigma_{\tilde{V}}} z;\rho) \nonumber \\ 
&+&\Phi_{W,V}( \frac{c_1}{\sigma_{\tilde{W}}}-\frac{\tilde{\beta}_0}{\sigma_{\tilde{W}}}-\frac{\tilde{\beta}_1}{\sigma_{\tilde{W}}} z,\frac{d_0}{\sigma_{\tilde{V}}}
-\frac{\tilde{\gamma_0}}{\sigma_{\tilde{V}}}-\frac{\tilde{\gamma_1}}{\sigma_{\tilde{V}}} z;\rho) \nonumber \\
&\equiv& \Phi^{1}_{c_{1},d_{1},d_{0}}, \label{IVTEQ1e} \\
P(c_1 \leq Y, D \leq d_0| Z=z) &=&  \Phi_{V}(\frac{d_0}{\sigma_{\tilde{V}}}
-\frac{\tilde{\gamma_0}}{\sigma_{\tilde{V}}}-\frac{\tilde{\gamma_1}}{\sigma_{\tilde{V}}} z) \nonumber \\
&-&\Phi_{W,V}( \frac{c_1}{\sigma_{\tilde{W}}}-\frac{\tilde{\beta}_0}{\sigma_{\tilde{W}}}-\frac{\tilde{\beta}_1}{\sigma_{\tilde{W}}} z,\frac{d_0}{\sigma_{\tilde{V}}}
-\frac{\tilde{\gamma_0}}{\sigma_{\tilde{V}}}-\frac{\tilde{\gamma_1}}{\sigma_{\tilde{V}}} z;\rho) \nonumber \\
&\equiv& \Phi^{1}_{c_{1},d_0},\label{IVTEQ1f}
\end{eqnarray}
\begin{eqnarray}
P(c_1\leq Y, D \geq d_1| Z=z) &=&1-\Phi_{W}( \frac{c_1}{\sigma_{\tilde{W}}}-\frac{\tilde{\beta}_0}{\sigma_{\tilde{W}}}-\frac{\tilde{\beta}_1}{\sigma_{\tilde{W}}} z)-\Phi_{V}(\frac{d_1}{\sigma_{\tilde{V}}}-\frac{\tilde{\gamma_0}}{\sigma_{\tilde{V}}}-\frac{\tilde{\gamma_1}}{\sigma_{\tilde{V}}} z) \nonumber \\
&+& \Phi_{W,V}( \frac{c_1}{\sigma_{\tilde{W}}}-\frac{\tilde{\beta}_0}{\sigma_{\tilde{W}}}-\frac{\tilde{\beta}_1}{\sigma_{\tilde{W}}} z,\frac{d_1}{\sigma_{\tilde{V}}}-\frac{\tilde{\gamma_0}}{\sigma_{\tilde{V}}}-\frac{\tilde{\gamma_1}}{\sigma_{\tilde{V}}} z;\rho) \nonumber \\
&\equiv& \Phi^{1}_{c_{1},d_{1}}.    \label{IVTEQ1c}
\end{eqnarray}
where $\Phi_{W,V}(w,v;\rho)$ denotes the joint c.d.f. of $(W,V)$, a standard bivariate normal with coefficient of correlation $\rho$. Similarly, considering the case that the observed $Y$ equals zero, 
\begin{eqnarray}\label{IVTEQ2}
P(Y=0, d_0 \leq D \leq d_1| Z=z)&=&\Phi_{W,V}( -\frac{\tilde{\beta}_0}{\sigma_{\tilde{W}}}-\frac{\tilde{\beta}_1}{\sigma_{\tilde{W}}} z,\frac{d_1}{\sigma_{\tilde{V}}}-\frac{\tilde{\gamma_0}}{\sigma_{\tilde{V}}}-\frac{\tilde{\gamma_1}}{\sigma_{\tilde{V}}} z;\rho)\nonumber \\ 
&-&\Phi_{W,V}(-\frac{\tilde{\beta}_0}{\sigma_{\tilde{W}}}-\frac{\tilde{\beta}_1}{\sigma_{\tilde{W}}} z,\frac{d_0}{\sigma_{\tilde{V}}}-\frac{\tilde{\gamma_0}}{\sigma_{\tilde{V}}}-\frac{\tilde{\gamma_1}}{\sigma_{\tilde{V}}} z;\rho)\equiv \Phi^{2}_{d_{1},d_{0}}, \label{IVTEQ2a} \\
P(Y=0,  D \leq d_0| Z=z)&=&\Phi_{W,V}( -\frac{\tilde{\beta}_0}{\sigma_{\tilde{W}}}-\frac{\tilde{\beta}_1}{\sigma_{\tilde{W}}} z,\frac{d_0}{\sigma_{\tilde{V}}}-\frac{\tilde{\gamma_0}}{\sigma_{\tilde{V}}}-\frac{\tilde{\gamma_1}}{\sigma_{\tilde{V}}} z;\rho)\equiv \Phi^{2}_{d_0}, \label{IVTEQ2b} \\
P(Y=0,  D \geq d_1| Z=z)&=&\Phi_{W}( -\frac{\tilde{\beta}_0}{\sigma_{\tilde{W}}}-\frac{\tilde{\beta}_1}{\sigma_{\tilde{W}}} z)-\Phi_{W,V}(-\frac{\tilde{\beta}_0}{\sigma_{\tilde{W}}}-\frac{\tilde{\beta}_1}{\sigma_{\tilde{W}}} z,\frac{d_1}{\sigma_{\tilde{V}}}-\frac{\tilde{\gamma_0}}{\sigma_{\tilde{V}}}-\frac{\tilde{\gamma_1}}{\sigma_{\tilde{V}}} z;\rho) \nonumber \\
&\equiv& \Phi^{2}_{d_{1}}. \label{IVTEQ3}
\end{eqnarray}
Then, the equalities (\ref{IVTEQ1})-(\ref{IVTEQ3}) are sharp testable implications for the validity of the instrumental variable Tobit Model in equations (\ref{IVT0}) coupled with assumptions \ref{IVT2}-\ref{IVT3}.  We collect these results more formally in Theorem \ref{Rem:SharpIV}.
\begin{theorem}\label{Rem:SharpIV}
Suppose that the IV Tobit model (\ref{IVT0}) or its reduced form representation (\ref{IVT1}) along with assumptions \ref{IVT2}-\ref{IVT3} hold. Then, the parameters $\alpha_{0}$, $\alpha_{1}$, $\gamma_{0}$, $\gamma_{1}$, $\sigma^{2}_{U}$, $\rho_{UV}$ are identified, and equalities (\ref{IVTEQ1})-(\ref{IVTEQ3}) must hold for all   $c,d, z \in \mathcal{Y} \times \mathcal{D} \times \mathcal{Z}$. Furthermore, these equalities are sharp, that is, whenever they hold, it is possible to construct a vector of $(\tilde{Y},\tilde{D},\tilde{U},\tilde{V},Z)$ or equivalently  $(\tilde{Y},\tilde{D},\tilde{W},\tilde{V},Z)$ that satisfies model \ref{IVT0} and \ref{IVT1}, Assumptions \ref{IVT2} and \ref{IVT3}, and induces the observed distribution on the data $(Y,D,Z)$.
\end{theorem}
 The proof for the equalities follows from the discussion above, and further details about sharpness are presented in Appendix \ref{app1}. They serve as the basis for the test procedure described in Section \ref{S4}.

\begin{proposition}[Non-learnability]\label{Rem:NLIV}
The testable implications and sharpness discussed above show that the IV-Tobit model is generally refutable. However, the model is non-verifiable. The demonstration follows a similar logic as in the classic Tobit case. See Appendix \ref{ApLearn} for the proof.
\end{proposition}
 Similarly to the discussion in Proposition \ref{RemEx}, we conjecture that the approaches proposed in Theorem \ref{Rem:SharpIV} can be adapted to models in which the researcher is willing to assume a known joint distribution for the latent errors, $(U,V)$, replacing the normal distribution, as long as the model is identified.

\section{Testing procedure} \label{S4}
To test the sharp equalities, we rewrite each of these equalities as two inequalities \citep{mourifie_testing_2017, acerenza_testing_2023}. 
For a concrete example,  we first note that  $$P(c_1 \leq Y|D=d) =1-\Phi\left(\frac{c_1}{\sigma}-\frac{\tilde{\alpha}_0}{\sigma}-\frac{\tilde{\alpha}_1}{\sigma} d\right) \iff E\left[1\{c_1 \leq Y \}-1+\Phi\left(\frac{c_1}{\sigma}-\frac{\tilde{\alpha}_0}{\sigma}-\frac{\tilde{\alpha}_1}{\sigma} D\right)|D\right]=0.$$
We rewrite each of these moment equalities implied by the restrictions on the empirical distribution as moment inequalities $$E\left[1\{c_1 \leq Y \}-1+\Phi\left(\frac{c_1}{\sigma}-\frac{\tilde{\alpha}_0}{\sigma}-\frac{\tilde{\alpha}_1}{\sigma}D\right)|D\right]\leq 0, E\left[-1\{c_1 \leq Y \}+1-\Phi\left(\frac{c_1}{\sigma}-\frac{\tilde{\alpha}_0}{\sigma}-\frac{\tilde{\alpha}_1}{\sigma} D\right)|D\right]\leq 0.$$ In doing this for all the restrictions on the empirical distribution, we can implement a test relying on existing intersection bounds inferential methods such as \cite{chernozhukov_intersection_2013}, which is specifically suited to test conditional moment inequalities.

Note that the equalities for the classic Tobit hold for any pair of constants $(c_{0},c_{1})$, and the ones from the IV Tobit hold for pairs of $(c_{0},c_{1})$ and $(d_{0},d_{1})$. We propose a partition of $\mathcal{Y}\times\mathcal{D}$ to test sufficient conditions of these sharp equalities. Rejection of any of the null hypotheses that the equalities hold at these particular levels implies violations of the model.
\subsection*{Classic Tobit model}\label{S41}
The sharp testable equalities for every $c_{0},c_{1} \in \mathcal{Y}$ are given by equations (\ref{CTEQ2}) and (\ref{CTEQ1}).
For a partition of the support of $Y$ into $K$ arbitrary chosen sets $C_k=(0,c_k)$ such that $C_k \in C_{k+1}$,  the following set of sufficient inequalities are, for some chosen values of $c_{k},c_{k+1} \in \mathcal{Y}$, related to the components of equations (\ref{CTEQ2})-(\ref{CTEQ1}).

The formulation of the inequalities considered will depend on each partition's location on the support of the outcome variable $Y$. Let $c_{1}=0$ and $W_{k}$ be
\begin{eqnarray*}
W_{k}=\left\{\begin{array}{cc}
     1\{Y=0\}-(1-\Phi(\frac{\tilde{\alpha}_0}{\sigma}+\frac{\tilde{\alpha}_1}{\sigma} D)), & \text{ if }k=0\\
     1\{c_{k} < Y\leq c_{k+1} \}-\Phi\left(\frac{c_{k+1}}{\sigma}-\frac{\tilde{\alpha}_0}{\sigma}-\frac{\tilde{\alpha}_1}{\sigma}D\right)+\Phi\left(\frac{c_{k}}{\sigma}-\frac{\tilde{\alpha}_0}{\sigma}-\frac{\tilde{\alpha}_1}{\sigma} D\right), & \text{ if } 1\leq k<K \\
     1\{c_{K} < Y\}-(1-\Phi\left(\frac{c_{K}}{\sigma}-\frac{\tilde{\alpha}_0}{\sigma}-\frac{\tilde{\alpha}_1}{\sigma} D\right)), & \text{ if }k=K\\
\end{array}
  \right.
\end{eqnarray*}
The intersection bounds framework considers the following $2(K+1)$ inequalities 
\begin{eqnarray*}
    \sup_{d}E[W_{k}|D=d]&\leq& 0\\
    \sup_{d}E[-W_{k}|D=d]&\leq& 0, \text{ for }k=0,\dots,K
\end{eqnarray*}
We can write more compactly as 
\begin{equation}\label{CLRT}
\max_{k}\sup_D\theta_k (D) \leq 0
\end{equation}
where $\theta_k(D)$ collects all the inequalities being tested. The decision rule for the test is given by \cite{chernozhukov_intersection_2013}, we reject $H_{0}$ if 
\begin{align}\label{Teststatistic}
\hat{\theta}_{1-\alpha}\equiv \max_{k}\sup_D\left\{\hat{\theta}(D,k)-\kappa_{1-\alpha}\hat{s}(D,k)\right\} > 0,
\end{align}
where $\hat{\theta}(D,k)$ is a nonparametric estimator for $\theta_{k}(D)$, $\hat{s}(D,k)$ its standard error, and $\kappa_{1-\alpha}$ is a critical value at the significance level $\alpha$. 
\subsection*{IV Tobit model}\label{S42}
For the instrumental variable Tobit model, the continuous support for both the outcome and treatment poses challenges to the implementation of the test.\footnote{Note that a discrete treatment can also be accommodated as an intermediate step between the current derivation and the derivations from \cite{acerenza_testing_2023}.} The sharp testable equalities for every $c_{0},c_{1} \in \mathcal{Y}$ and $d_{0},d_{1} \in \mathcal{D}$ are given by equations (\ref{IVTEQ1a})-(\ref{IVTEQ3}).

Consider a partition of the support of $Y$ into $K$ arbitrary chosen sets $C_k=(0,c_k)$ such that $C_k \in C_{k+1}$ and of the support of $D$ into $Q$ arbitrary chosen sets $D_q=(0,d_q)$ such that $D_q \in D_{q+1}$,  the following set of sufficient inequalities are related to the components of (\ref{IVTEQ1a})-(\ref{IVTEQ3}). Analogous to the classic Tobit case, the formulation of the inequalities considered will depend on each partition's location on the joint support of the outcome and treatment variables.
Let $W_{kq}$ be given by,
\begin{eqnarray*}
W_{kq}=\left\{\begin{array}{cc}
     1\{ Y=0\}1\{D \leq d_0\}-\Phi^{2}_{d_{0}}, & \text{ if }k=0,q=0\\
     1\{Y=0\}1\{d_{q}\leq D \leq d_{q+1}\}-\Phi^{2}_{d_{q},d_{q+1}}, & \text{ if }k=0, 1\leq q<Q\\
     1\{ Y=0\}1\{D \geq d_Q\}-\Phi^{2}_{d_{Q}}, & \text{ if }k=0,q=Q\\
     1\{c_{k} \leq Y \leq c_{k+1}\}1\{D \leq d_0\}-\Phi^{1}_{c_{k+1},c_{k},d_{0}}, & \text{ if } 1\leq k<K,q=0\\
     1\{c_{k} \leq Y \leq c_{k+1}\}1\{d_{q} \leq D \leq d_{q+1}\}-\Phi^{1}_{c_{k+1},c_{k},d_{q+1},d_q}, & \text{ if } 1\leq k<K, 1\leq q<Q\\
     1\{c_{k} \leq Y \leq c_{k+1}\}1\{D \geq d_Q\}-\Phi^{1}_{c_{k+1},c_{k},d_Q}, & \text{ if } 1\leq k<K, q=Q\\
     1\{Y \geq c_{K}\}1\{D \leq d_{0}\}-\Phi^{1}_{c_{K},d_{0}}, & \text{ if } k=K,q=0\\
     1\{Y \geq c_{K}\}1\{d_{q} \leq D \leq d_{q+1}\}-\Phi^{1}_{c_{K},d_{q+1},d_{q}}, & \text{ if } k=K, 1\leq q<Q\\
     1\{Y \geq c_{K}\}1\{D \geq d_Q\}-\Phi^{1}_{c_{K},d_{Q}}, & \text{ if } k=K, q=Q.\\
\end{array}
  \right.
\end{eqnarray*}
Where we used the simplifying notation defined in equations (\ref{IVTEQ1a})-(\ref{IVTEQ3}) for each partition of the support for $Y$ and $D$.
The intersection bounds framework considers the following $2(K+1)(Q+1)$ inequalities 
\begin{eqnarray*}
    \sup_{z}E[W_{kq}|Z=z]&\leq& 0\\
    \sup_{z}E[-W_{kq}|Z=z]&\leq& 0, \text{ for }k=0,\dots,K;q=0,\dots,Q.
\end{eqnarray*}
We can write more compactly as 
\begin{eqnarray}\label{CLRTIV}
\max_{k,q}\sup_Z\theta_kq (Z) \leq 0
\end{eqnarray}
where $\theta_kq(Z)$ collects all the inequalities being tested. The decision rule for the test is given by \cite{chernozhukov_intersection_2013}, we reject $H_{0}$ if 
\begin{align}\label{TeststatisticIV}
\hat{\theta}_{1-\alpha}\equiv \max_{k,q}\sup_Z\left\{\hat{\theta}(Z,k,q)-\kappa_{1-\alpha}\hat{s}(D,k,q)\right\} > 0,
\end{align}
where $\hat{\theta}(Z,k,q)$ is a nonparametric estimator for $\theta_{kq}(Z)$, $\hat{s}(D,k,q)$ its standard error, and $\kappa_{1-\alpha}$ is a critical value at the significance level $\alpha$. 

To implement the test within the \cite{chernozhukov_intersection_2013} intersection bounds inferential method, we use the CLR Stata package described in \cite{chernozhukov_implementing_2015}. The parameters in the relevant model, for example, $\beta_{0}, \beta_{1}, \gamma_{0}, \gamma_{1}, \rho$, are replaced by their maximum likelihood estimators (MLE), and asymptotic validity of this ``plug-in'' test follows from the argument described by \cite[Appendix B]{acerenza_testing_2023}. Some additional details on the test implementation are discussed in Section \ref{S5}.
\begin{remark}
Intuitively, the testable conditions derived above consider whether the empirical conditional distribution of the observed outcome variable — in both the mass accumulation and non-truncated parts of the support of $Y$ — is consistent with a random variable(s) following the (bivariate) normal distribution for different sections of the distribution and values of the independent instrument, $Z$.
\par 
 The proposed test procedures are intended to detect violations of the model due to:
 \par 
1. Misspecification of the latent structure that makes the coefficient estimates biased as estimates of the true coefficients of $Y^*$;
\par 
2. Violations arising from the empirical distribution of $Y$ being inconsistent with the implied distributions from the parametric structure (that is if the proportion of residuals in different parts of its support deviate from the normality assumptions);
\par
3.  Violations due to the empirical distributions of the residuals differing from the implied distributions in certain values of the treatment (Classic Tobit) or instrument (IV Tobit), which indicate violations of the exogeneity of treatment or instrument \citep{kedagni_generalized_2020}.
\end{remark}
\begin{remark}
Alternative approaches to test the classic and IV Tobit could be considered. For example, an intuitive approach would be calculate the residuals $\hat{U}=Y-\hat{\alpha}_{0}-\hat{\alpha}_{1}D$ and compare its distribution to the (truncated) std. normal through one of the usual normality tests in the literature.\footnote{We received this suggestion in several seminars and from anonymous referees, to whom we are thankful.} One of the challenges in doing so is that the latent variable is recoverable (as estimated by the residuals) only if $Y=\alpha_{0}+\alpha_{1}D+U>0$, while for all observations for which $Y=0$ the only information available is that $\alpha_{0}+\alpha_{1}D\leq-U$. Thus, in our case, all observations that are “at the corner” would only give us information about U being below a certain truncation value, which depends on $D$ (and any other covariates included in the model, see below), which is a non-pivotal
quantity and is not associated with any well-established distributional test. A second approach would follow \cite{li_nonparametric_2023}  by testing for correct parametric functional forms for $E[1\{c_0\leq Y \leq c_1 \}|D]$ against $\Phi(c_1-\alpha_0-\alpha_1 D)-\Phi(c_0-\alpha_0-\alpha_1 D)$, which is similar in spirit to our test if one incorporates all the equalities proposed in Theorem \ref{Rem:SharpClassic}-\ref{Rem:SharpIV}. That approach would also require partitioning the support of the variables being considered. A nonparametric test statistic could be used, following  \cite{li_nonparametric_2023}, the details of which are beyond the scope of this manuscript. 
Both approaches would present difficulties of implementation at least as important as the ones faced by proposed test based in the inferential approach by \cite{chernozhukov_implementing_2015} discussed above. While recognizing that CLR may lead to a conservative test, the relatively easy implementation of the test using available statistical packages is an attractive feature. Further research on different approaches and their properties could lead to valuable alternatives.
\end{remark}
\subsection*{Including Covariates}\label{S43}
The proposed procedure can be extended to include exogenous covariates, $X$, within the linear index model in \ref{IVT0}. Then, the testable equalities for classic and IV Tobit models can be derived with the additional conditioning on $X$.
The test with covariates could be implemented by generating a partition of the covariate space, say in $J$ grids, similar to the partition of the exogenous variable used in Section \ref{S4}. For every grid in the partition, one computes the test statistic \ref{Teststatistic} or \ref{TeststatisticIV} for the classic or the IV-tobit, respectively. Then obtain critical values that account for multiple testing via a Bonferroni correction. In particular, for every grid, set the critical value at the significance level $\frac{\alpha}{J}$, namely $\kappa_{1-\frac{\alpha}{J}}$. \footnote{A theoretically interesting approach would partition the joint support of all the exogenous variables and compute the test statistics across all parts of the grid. However, this approach would entail significant implementation challenges since the statistical package for \cite{chernozhukov_implementing_2015} allows for only one exogenous covariate.}
This procedure can be cumbersome when there are many covariates or when they are continuous.

Another route that is less computationally intensive follows \cite{acerenza_testing_2023}. We illustrate it for the $IV$-tobit case, and a similar logic could be applied to the classic Tobit case. Thus, with covariates added to the linear index function, we have,
\begin{eqnarray}\label{IVTCov2}
\left\{ \begin{array}{lcl}
     Y &=& \max(0,Y^*) \\
     Y^* &=& \tilde{\beta}_0+\tilde{\beta}_1 Z+\tilde{\beta}_2X+\tilde{W}  \\
     D&=& \tilde{\gamma}_0+\tilde{\gamma}_1 Z+\tilde{\gamma}_2 X+\tilde{V}
     \end{array} \right.
\end{eqnarray}
We extend Assumption \ref{IVT2} to formalize covariates' exogeneity:
 \begin{assumption}[full independence] \label{IVT2full}
$(Z,X)\ \indep\ (\tilde{W},\tilde{V})$.
\end{assumption} 
Under Assumptions \ref{IVT3}-\ref{IVT2full}, by a similar argument to that used in Section \ref{S9}, any given generic testable implication becomes 
\begin{eqnarray}\label{EQcovariates}
P(c_0\leq Y \leq c_1, d_0 \leq D \leq d_1| Z=z, &X&=x)= \Phi_{W,V}\left( \frac{c_1}{\sigma_{\tilde{W}}}-\frac{\tilde{\beta}_1}{\sigma_{\tilde{W}}} z-\frac{\tilde{\beta}_2}{\sigma_{\tilde{W}}}x,\frac{d_1}{\sigma_{\tilde{V}}}-\frac{\tilde{\gamma_0}}{\sigma_{\tilde{V}}}-\frac{\tilde{\gamma_1}}{\sigma_{\tilde{V}}} z-\frac{\tilde{\gamma_2}}{\sigma_{\tilde{V}}}x;\rho\right) \nonumber\\
&-&\Phi_{W,V}\left(\frac{c_0}{\sigma_{\tilde{W}}}-\frac{\tilde{\beta}_0}{\sigma_{\tilde{W}}}-\frac{\tilde{\beta}_1}{\sigma_{\tilde{W}}} z-\frac{\tilde{\beta}_2}{\sigma_{\tilde{W}}}x,\frac{d_1}{\sigma_{\tilde{V}}}-\frac{\tilde{\gamma_0}}{\sigma_{\tilde{V}}}-\frac{\tilde{\gamma_1}}{\sigma_{\tilde{V}}} z-\frac{\tilde{\gamma_2}}{\sigma_{\tilde{V}}}x;\rho\right) \nonumber \\ 
&-& \Phi_{W,V}\left(\frac{c_1}{\sigma_{\tilde{W}}}-\frac{\tilde{\beta}_1}{\sigma_{\tilde{W}}} z-\frac{\tilde{\beta}_2}{\sigma_{\tilde{W}}}x,\frac{d_0}{\sigma_{\tilde{V}}}-\frac{\tilde{\gamma_0}}{\sigma_{\tilde{V}}}-\frac{\tilde{\gamma_1}}{\sigma_{\tilde{V}}} z-\frac{\tilde{\gamma_2}}{\sigma_{\tilde{V}}}x;\rho\right) \nonumber \\
&+&\Phi_{W,V}\left(\frac{c_0}{\sigma_{\tilde{W}}}-\frac{\tilde{\beta}_1}{\sigma_{\tilde{W}}} z-\frac{\tilde{\beta}_2}{\sigma_{\tilde{W}}}x,\frac{d_0}{\sigma_{\tilde{V}}}-\frac{\tilde{\gamma_0}}{\sigma_{\tilde{V}}}-\frac{\tilde{\gamma_1}}{\sigma_{\tilde{V}}} z-\frac{\tilde{\gamma_2}}{\sigma_{\tilde{V}}}x;\rho\right)
\end{eqnarray}
for all $z \in \mathcal Z$ and $x \in \mathcal X$, where $x$ can be a vector.  Equivalent conditions for the case that $Y=0$ and other parts of the support of $Y$ and $D$ can be similarly obtained.

\subsubsection*{Implementation} 
Since the test is nonparametric, we face challenges when $X$ is high-dimensional, especially with continuous covariates. Another operational limitation is that the Stata's clrbound package only allows for one conditioning variable at a time. For those practical reasons, we propose implementing a weaker, non-sharp, version of the testable equalities. Continuing with the example of Equality \ref{EQcovariates}, we can integrate over the covariates $X$ (or alternatively, $Z$), taking advantage of the fact that $\mathbb E[W\vert Z=z, X=x]=0$ implies $\mathbb E[W\vert Z=z]=0$ and $\mathbb E[W\vert X=x]=0$ for all random variables $W$.

Implementation becomes very similar to the IV-Tobit discussed above by redefining the simplifying notation in equations (\ref{IVTEQ1a})-(\ref{IVTEQ3}), with the only difference being the inclusion of $X$ on the linear indexes in $\Phi_{W}(\cdot)$ and $\Phi_{W,V}(\cdot,\cdot,\cdot)$. For example, $\Phi^{1}_{c_{1},c_{0},d_{1},d_{0}}$ is given by equation \eqref{EQcovariates} and similarly for the other terms. Then, compute the new $W_{kq}$ in the same manner as in Section \ref{S42}, and the intersection bounds framework considers the similar $2(K+1)(Q+1)$ inequalities, based on the partition of the support of $Y$ and $D$. 
\begin{eqnarray*}
    \sup_{z}E[W_{kq}|Z=z]&\leq& 0\\
    \sup_{z}E[-W_{kq}|Z=z]&\leq& 0, \text{ for }k=0,\dots,K;q=0,\dots,Q.
\end{eqnarray*}

This is the test procedure that we implement for the empirical application in Section \ref{S7b}.

Under Assumption \ref{IVT2full}, one could base the test on conditioning on a particular covariate $X_{C}$ instead of $Z$ by integrating the sharp Equality (\ref{EQcovariates}) over $Z$ and then implementing the intersection bounds procedure using $X_{C}$ as the sole conditioning variable.\footnote{Assumption \ref{IVT2full} constraints the relationship between the covariates and the latent error terms in such a way that one could combine the information on all $X$ and $Z$ in an index and construct non-sharp testable equalities that would hold conditional on this index. For example, having $\sup_{t}E[W_{kq}|Z+X_1+...+X_C=t]= 0$, for $k=0,\dots,K;q=0,\dots,Q$.}

\section{Simulations} \label{S5}

In this section, we provide simulation exercises for the proposed tests. The testing procedure described relies on testing sharp equalities that should hold for any arbitrary partition of the outcome, treatment and instrument support. When both the outcome and treatment are continuous, evaluating all possible equalities is technically challenging. We focus on a non-sharp set of the equalities by evaluating them at different grid partitions of their support, in a similar spirit to \cite{honore2020selection} which is particularly well suited for continuous supports. Naturally, this choice makes the test less powerful as we don't consider the continuum of equalities derived in Section \ref{S3}, but is justified by the ease of implementation of the procedure based on intersection bounds and the performance of the test on the simulations below.

For the simulations related to the classic Tobit, we partition the support of the observed outcome variable into the accumulation point ($Y=0$) and four quartiles on the (untruncated) positive range, while for the IV-Tobit case we also partition the support of the treatment variable ($D$) to create a grid based on both the outcome and treatment.\footnote{Note that if the treatment variable was discrete, the respective grid points could be set naturally to the possible countable values $D$ takes.}
The choice of the number and location of the partitions/evaluation points balances the implementation computational requirements, data availability for different parts of the joint support of the outcome, treatment and exogenous instrumental variable. For the procedure in Section \ref{S4} to be feasible, we must have data on both the outcome and the exogenous variables within each partition. Using the empirical quantiles of the non-truncated outcomes to determine the partitions guarantees a reasonable number of observations for each grid part. Larger sample sizes might allow finer partitions for the outcome support. Still, the added computational requirements created by an increased number of equalities being checked, coupled with the larger datasets, can substantially increase computing time.\footnote{The simulations presented in Section \ref{S5} have limited sample sizes and a relatively small number of equalities being tested due to the long-running time and computational constraints when repeating the test procedure thousands of times.}

When implementing the intersection bounds in STATA using the package clrbound \cite{chernozhukov_implementing_2015}, the researcher must determine the range of values of the exogenous variable for which each equality will be evaluated. To guarantee the feasibility of the procedure, we adjust the evaluation points for $D$ ($Z$) to the first and $99^{th}$ percentiles of the exogenous variable in each partition of the support for the outcome $Y$.\footnote{See details on implementation on the simulation replication STATA code.}

In evaluating the finite sample performance of the proposed tests, we consider a continuous treatment $D$ following the data generating process described in Equation \ref{eq:SizeDGP}.

  \begin{eqnarray}
 \label{eq:SizeDGP}
 \left\{ \begin{array}{lcl}
      Y &=& \max\{Y^*,0\}\\
      Y^* &=& D+U \\ 
      D&=&2Z-V   \\ 
    \begin{pmatrix}
U \\
V \\
Z 
\end{pmatrix} &\sim& \mathcal N(\mathbf{0}_{3}, \mathbf{\Sigma}) \\
\mathbf{\Sigma}&=&\begin{pmatrix}
1 & \rho_{uv} & \rho_{uz}\\
 \rho_{uv} & 1 & \rho_{vz} \\
 \rho_{uz} &  \rho_{vz} & 1
\end{pmatrix} \\
\rho_{uz}&=&\rho_{vz}=0 \\
\rho_{uv}&=&\rho
    \end{array} \right.
\end{eqnarray}

 Where $\mathbf{0}_{p}$ is a $p \times 1$ vector of zeroes. The parameter $\rho$ determines the intensity of dependence between the latent variables jointly determining the treatment and outcome. Under the treatment exogeneity condition described in Assumption \ref{CT2}, $\rho=0$. Table \ref{tab:Size_Test} presents the empirical test sizes for the Classic Tobit test this scenario, for different significance levels $\alpha$. The results indicate that while the test over rejects the null hypothesis for small to mid-sized samples, the test's empirical coverage approaches its desired nominal benchmark as samples larger than 5,000 are used.

\begin{table}[!htbp]
\caption{Classic Tobit Test Size}
\label{tab:Size_Test}
\begin{tabular}{l|ccc}
N &$\alpha=10\%$&$\alpha=5\%$  &$\alpha=1\%$\\  \hline
1000	&18.84$\%$ &	14.23$\%$ &	9.62$\%$ \\
2000	&16.00$\%$ &	11.00$\%$ &	5.00$\%$ \\
5000	&10.82$\%$ &	6.21$\%$ &	3.01$\%$ \\
8000	&10.00$\%$ &	6.40$\%$ &	3.20$\%$ \\
10000	&8.80$\%$ &	5.40 $\%$ &	1.80$\%$ 
 \\ \hline
\end{tabular}
\begin{center}
\scriptsize{Based on $500$ replications.}
\end{center}
\end{table}

To consider the test's performance under violations of the exogeneity assumption, we modify the DGP in Equation \ref{eq:SizeDGP} with different values for $\rho$, reflecting various degrees of treatment endogeneity. Larger values of $\rho$ produce more acute violations of the null hypothesis. Table \ref{tab:PowerRho_Test} presents the results for the Classic Tobit test. As expected, the power of the test increases with larger $\rho$ and bigger sample sizes.

\begin{table}[!htbp] 
\caption{Classic Tobit test power for violations in exogeneity}
\label{tab:PowerRho_Test}
\begin{tabular}{ll|ccc}
    $N$ & $\rho$ & $\alpha=10\%$ & $\alpha=5\%$ & $\alpha=1\%$ \\  \hline
        5000 & 0.10  & 8.60\% & 6.00\% & 3.20\% \\ 
        ~ & 0.50 & 16.03\% & 10.82\% & 5.21\% \\ 
        ~ & 0.75 & 36.00\% & 27.80\% & 14.20\% \\ 
        ~ & 0.80  & 42.00\% & 30.60\% & 16.80\% \\ 
        ~ & 0.90  & 62.40\% & 48.40\% & 27.60\% \\ 
        ~ & 0.95  & 71.34\% & 59.32\% & 35.87\% \\  \hline
        8000 & 0.10  & 8.40\% & 5.00\% & 2.00\% \\ 
        ~ & 0.50 &  17.43\% & 12.22\% & 5.61\% \\ 
        ~ & 0.75  & 45.58\% & 36.14\% & 20.08\% \\ 
        ~ & 0.80 & 53.80\% & 39.80\% & 18.00\% \\ 
        ~ & 0.90 & 79.00\% & 65.40\% & 36.40\% \\ 
        ~ & 0.95 & 85.57\% & 75.15\% & 49.70\% \\  \hline
        10000 & 0.10 & 10.04\% & 7.43\% & 2.41\% \\ 
        ~ & 0.50 &  15.80\% & 10.00\% & 3.80\% \\ 
        ~ & 0.75 &  44.80\% & 32.80\% & 16.00\% \\ 
        ~ & 0.80 &  56.60\% & 41.00\% & 21.60\% \\ 
        ~ & 0.90 & 79.60\% & 66.60\% & 36.60\% \\ 
        ~ & 0.95 &  90.56\% & 79.52\% & 51.61\% \\ \hline
\end{tabular}
\begin{center}
\scriptsize{Based on $500$ replications.}
\end{center}
\end{table}

To consider a violation with respect to the violation of the normality of errors. We consider a  modified DGP from Equation \ref{eq:SizeDGP}, by seting $(V, Z)^\top \sim \mathcal{N}(\mathbf{0}_2, \mathbf{I}_2)$ and $U\sim F(\cdot)$. Table \ref{tab:PowerDist_Test} presents the power results  when $F(\cdot)$ is uniform, lognormal, and t-student distributions. As expected, the rejection rate for the t-student test with 80 degrees of freedom is close to the nominal size of the test, since it represents a very mild violation of the normality assumption. In the other three cases, which deviate significantly from the normality assumption, the test rejects the null hypothesis in most instances. This example demonstrates that the proposed test is effective in identifying violations of the distributional assumptions in the classic Tobit model. 

\begin{table}[!htbp] 
\caption{Classic Tobit test power for violations in the error structure}
\label{tab:PowerDist_Test}
\begin{tabular}{ll|ccc}
    $N$ & $U$ & $\alpha=10\%$ & $\alpha=5\%$ & $\alpha=1\%$ \\  \hline
5000 & t-student (df=80) & 7.21\% & 4.01\% & 1.40\% \\
& t-student (df=5) & 92.20\% & 81.80\% & 48.80\% \\
& LogNormal & 100.00\% & 100.00\% & 100.00\% \\
& Uniform & 100.00\% & 100.00\% & 100.00\% \\ \hline
8000  & t-student (df=80)  & 7.00\% & 4.80\% & 1.20\% \\
& t-student (df=5)  & 99.20\% & 97.80\% & 87.58\% \\
& LogNormal & 100.00\% & 100.00\% & 100.00\% \\
& Uniform & 100.00\% & 100.00\% & 100.00\% \\ \hline
10000 & t-student (df=80)  & 5.42\% & 2.41\% & 1.00\% \\
& t-student (df=5) & 99.20\% & 98.40\% & 90.58\% \\
& LogNormal & 100.00\% & 100.00\% & 100.00\% \\ 
& Uniform & 100.00\% & 100.00\% & 100.00\% \\
\hline
\end{tabular}
\begin{center}
\scriptsize{Based on $500$ replications.}
\end{center}
\end{table}

Naturally, researchers concerned about treatment endogeneity should consider the IV-Tobit model and implement the test of its identifying assumptions proposed in Section \ref{S42}. Table \ref{tab:Size_Test_IV} presents the empirical coverage for the test of the IV-Tobit model for different levels of treatment endogeneity ($\rho=\{0, 0.5, 0.8\}$) for sample sizes 5,000 and 8,000. The test produces adequate empirical coverage, in line with the results for the classic Tobit test.  
\begin{table}[!htbp]
\caption{IV Tobit - Test Size}
\label{tab:Size_Test_IV}
\begin{tabular}{ll|ccc}
N & $\rho_{uv}$ &$\alpha=10\%$&$\alpha=5\%$  &$\alpha=1\%$\\  \hline
    1000  & 0     & 30.20\% & 25.60\% & 18.00\% \\
      & 0.5   & 34.80\% & 29.00\% & 20.00\% \\
      & 0.8   & 29.60\% & 23.40\% & 15.40\% \\
    \midrule
    2000  & 0     & 20.20\% & 15.60\% & 8.20\% \\
      & 0.5   & 21.40\% & 16.60\% & 9.00\% \\
      & 0.8   & 22.60\% & 15.20\% & 9.40\% \\
    \midrule
    5000  & 0     & 10.80\% & 7.00\% & 3.20\% \\
      & 0.5   & 11.40\% & 9.80\% & 3.40\% \\
      & 0.8   & 14.40\% & 9.60\% & 4.60\% \\
    \midrule
    8000  & 0     & 7.40\% & 3.80\% & 1.80\% \\
      & 0.5   & 9.20\% & 5.60\% & 1.80\% \\
      & 0.8   & 15.20\% & 10.00\% & 3.60\% \\
    \midrule
    10000 & 0     & 8.68\% & 5.21\% & 1.74\% \\
     & 0.5   & 10.00\% & 6.60\% & 1.80\% \\
     & 0.8   & 12.00\% & 8.60\% & 3.40\% \\
 \hline
\end{tabular}
\begin{center}
\scriptsize{Based on $500$ replications.}
\end{center}
\end{table}

\section{Relaxation of the assumptions}\label{S6}
When the test proposed in Section \ref{S4} rejects the null hypothesis of the model's validity, researchers must pursue alternative models and less restrictive assumptions to learn confidently about the parameters of interest.
\subsection{Alternative Approaches}
There is a vast literature on alternatives to the Tobit Model that can be implemented in the presence of censored dependent variables. Most approaches consider changes or relaxations of one of the two main assumptions associated with the model. The first assumption is the parametric distribution of the error terms and latent index form connecting treatment (and covariates) to the outcome. The second assumption is the exogeneity of the treatment or potential instrument. Here, we provide a non-exhaustive survey of existing work. 

\cite{cragg1971some} maintains the normality of the errors, linearity of the index and exogeneity but relaxes the way censoring occurs in comparison to the latent structure of the censored outcome. Specifically, while the latent outcome is still modelled by $Y^*=\alpha_0+\alpha_1 D+U$, they allow for the censoring to depend on a different linear index, $P(Y^*>0)=P(\gamma_0+\gamma_1 D+e)$, increasing the model's flexibility. 

\cite{POWELL1984303} relaxes the parametric structure of the errors while maintaining the latent linear index and treatment exogeneity, and estimates the parameters of interest by least absolute deviations. \cite{powell1986symmetrically} also maintains linearity of the latent index and treatment exogeneity, but relaxes normality by imposing symmetrical distributions to the latent errors, which leads to estimation by symmetrically censored least squares. \cite{newey1987efficient}  relaxes exogeneity of the treatment, relying on normality and an instrumental variable to identify the model, which is estimated by generalized least squares.

\cite{honore1994pairwise} relax linearity and mean independence of the unobservable with respect to the treatment to exploit the idea that, although $Y^*_i-\alpha_0-\alpha_1D_i$ is not mean-independent of $D_i$, one can trim any pair of residuals  $Y^*_i-\alpha_0-\alpha_1D_i$ and  $Y^*_j-\alpha_0-\alpha_1D_j$, and the trimmed residuals are independent and identically distributed conditional on  $D_i,D_j$. They estimate the model by identically censored least absolute deviations and identically censored least squares (ICLS). \cite{das2002estimators} estimates a model using symmetrically censored least squares that relaxes exogeneity of the treatment and normality of the errors. To achieve that they rely on instrumental variables, linearity of the mean of the structural error conditional on the reduced form error, and mean independence of the reduced form error.

\cite{blundell2007censored} proposes a control variable approach that relaxes exogeneity and normality but maintains the latent linear structure ($\alpha_0+\alpha_1 D+U$). Crucially, they impose that the distribution (or quantiles) of the latent error conditional on the treatment and instrument is only a function of the control variable $V=D-\pi(Z)$, which isolates the endogenous variation on the treatment.\footnote{This assumption is weaker than independence of all errors and instruments since it does not impose $V$ independent of $Z$ but is neither stronger nor weaker than independence of $U$ and $Z$, since it permits $Z$ to affect $U$ through $V$.} This allows them to estimate the effect of the treatment by censored quantile instrumental variable regression augmented by a control variable given by the quantiles of $U$ conditional on $V$ at the quantile of interest.  In a similar spirit, \cite{chernozhukov2015quantile} focuses on conditional quantile functions and flexible approaches to estimate the control variable in the first stage. 

Finally, \cite{chesher2023iv} provides partial identification results for a general alternative by relaxing the exogeneity of the treatment and instrument, linearity of the latent index and imposing no parametric structure of the error term. They characterize the identified set for the parameters of interest following the Generalized Instrumental Variables framework \citep{chesher2017generalized}, relying on the assumption that the relationship of $Y^*$ to treatment and errors is continuous and monotonic in the errors. Their approach uses the residual sets associated with the structure of the latent function and conditional probability of the error term given potential instruments.

\subsection{Partial identification under monotonicity}
In this subsection, we present an approach that partially identifies the effect of an endogenous treatment variable by replacing the normality and exogeneity assumptions with a monotonicity in treatment selection constraint. While less general than \cite{chesher2023iv}, this approach is easy to implement and could be useful to empirical researchers. 
\par 
Consider the model that maintains linearity (or a known structure of $Y^*$ up to a finite number of parameters), 
 \begin{eqnarray}\label{R1}
\left\{ \begin{array}{lcl}
     Y &=& \max(0,Y^*) \\
     Y^* &=& \alpha_0+\alpha_1 D+U
     \end{array} \right.
\end{eqnarray}

As an alternative to treatment exogeneity and normality, consider a constraint on the direction of the endogenous relationship between the treatment and the unobservables that affect the outcome.

\begin{assumption}[Monotone Treatment Selection - MTS]\label{R2}
Let $E(U|D=d,Y>0)\equiv \Gamma(d)$. Then, for  any $d>d^*$ we either have $\Gamma(d) < \Gamma(d^*)$ or  $\Gamma(d) > \Gamma(d^*)$.
\end{assumption}

Assumption \ref{R2} is common in the partial identification literature \citep{jiang2014monotone,manski1997monotone,manski2000amonotone, okumura2014concave}. In this context, we restrict the latent selection to be monotonic with respect to the treatment. This assumption is embedded in the classic Tobit model since the inverse mills ratio, $\lambda(\cdot)$, is monotonic (and decreasing) in the treatment variable. Furthermore, independence between $D$ and $U$ restricts the sign of the coefficient of the selection term directly, as the derivative of the inverse mills-ratio is $\lambda^{\prime} (\alpha_0+\alpha_1 D) \alpha_1$. Thus, without imposing independence or a parametric latent structure, we maintain a relevant property of the Tobit model that aids identification. Since it is not as restrictive as imposing a parametric structure and independence, we can only partially identify the parameter of interest.
\par 
Under Assumption \ref{R2} and the model described in equations (\ref{R1}), treatment and outcome are not independent. Note that,  
 \begin{eqnarray*}\label{R3}
      E(Y|D=d,Y>0) &=&  \alpha_0+\alpha_1 d+\Gamma(d)\\
      E(Y|D=d,Y>0)&-&\alpha_0-\alpha_1 d=\Gamma(d)
\end{eqnarray*}
Then, for any two $d>d^*$ we have, by Assumption \ref{R2}: 
 \begin{eqnarray}\label{R5}
 \begin{array}{lcl}
    \Gamma(d) &<&  \Gamma(d^*) \\  &\Leftrightarrow&\\
     E(Y|D=d,Y>0)-\alpha_0-\alpha_1 d &<&  E(Y|D=d^*,Y>0)-\alpha_0-\alpha_1 d^* \\ 
     &\Leftrightarrow& \\ 
     \alpha_1 &>& \frac{E(Y|D=d^*,Y>0)-E(Y|D=d,Y>0)}{d^*-d},
     \end{array}
\end{eqnarray}
which implies a lower bound on the parameter interest.
\par 
For a binary treatment $D\in \{0,1\}$ the lower bound is, intuitively, the difference in average outcomes between treated and untreated individuals away from the mass point at zero: 
 \begin{eqnarray}\label{R6}
 \begin{array}{lcl}
     \alpha_1 &>& E(Y|D=1,Y>0)-E(Y|D=0,Y>0).
     \end{array}
\end{eqnarray}
The bound can be more informative in the case of a continuous or multi-valued treatment. If $\Gamma(D)$ is differentiable we have:
 \begin{eqnarray}\label{R6cont}
     \alpha_1 &>& \frac{\partial E(Y|D=d,Y>0)}{\partial d} \text{, for all d.}
\end{eqnarray}
Since the inequality holds for any $d$ in the continuous case or for any $d,d^{*}$ for multi-valued discrete treatment, the linear index structure with constant parameters implies that tighter bounds for $\alpha_{1}$ are given by the maximum value of $\frac{\partial E(Y|D=d,Y>0)}{\partial d}$ across all possible points in the support for $D$. Analogous results with the inequalities reverted can be derived for any $d>d^*$, as we have $\Gamma(d)>\Gamma(d^*)$.

\par 
One-sided simple confidence regions can be computed based on these outer sets of the treatment effect.  One can estimate  $ E(Y|D=d,Y>0)$  using its sample analogs and rely on their asymptotic normality. For example, let the estimators be given by  $\widehat{E}(Y|D=d,Y>0)$. By the continuous mapping theorem, $ \frac{\widehat{E}(Y|D=d^*,Y>0)-\widehat{E}(Y|D=d,Y>0)}{d^*-d}$ is asymptotically normal. Thus, a one-sided confidence interval for $ \frac{E(Y|D=d^*,Y>0)-E(Y|D=d,Y>0)}{d^*-d}$ can be computed via bootstrap, which implies a conservative estimate for the lower bound for $\alpha_1$. 

A bootstrap procedure could be used for a nonparametric estimator of $\frac{\partial E(Y|D=d,Y>0)}{\partial d}$ using a local polynomial regression. In this case, the estimator is asymptotically normal and converges at a nonparametric rate that depends on the bandwidth $h$. Recent developments in \cite{calonico2018effect, calonico2022coverage} for estimation and optimal coverage error bandwidth and kernel selection methods that are nonparametric robust bias-corrected (RBC) can be used through their convenient implementation using the package nprobust \citep{calonico2019nprobust}.
Alternatively, since $\alpha_1 > sup_{d} \frac{\partial E(Y|D=d,Y>0)}{\partial d}$, one could consider obtaining confidence regions for $\alpha_1$ using a CLR approach similar to that described in Section \ref{S4} by considering nonparametric estimates of $\frac{\partial E(Y|D=d,Y>0)}{\partial d}$ or its discrete counterpart.\footnote{We thank an anonymous referee for this suggestion.}
\begin{remark}[On including covariates]\label{remcov}
 If we impose that $Y^* = \alpha_0+\alpha_1 D+\alpha_{2}X+U$ the procedure can include exogenous covariates by modifying Assumption \ref{R2} to hold conditional on $X$. Then, 
  \begin{eqnarray}
     \Gamma(d,X) &<&  \Gamma(d^*,X) \nonumber\\  
     &\Leftrightarrow&\nonumber\\
     E(Y|D=d,X,Y>0)-\alpha_0-\alpha_1 d-\alpha_2 X &<&  E(Y|D=d^*,X,Y>0)-\alpha_0-\alpha_1 d^*-\alpha_2 X \\ 
     &\Leftrightarrow&\nonumber \\ 
     \alpha_1 &>& \frac{E(Y|D=d^*,X,Y>0)-E(Y|D=d,X,Y>0)}{d^*-d}.\nonumber
 \end{eqnarray}
A similar argument holds when the treatment variable is continuous. Tighter bounds are achieved by considering the maximum value of $\frac{\partial E(Y|D=d,X=x,Y>0)}{\partial d}$ across all possible points in the support for $D$ and $X$. In practice, nonparametric estimates of these derivatives can be difficult, even for moderate numbers of covariates, and particularly challenging when multiple continuous covariates are in the conditioning set. The estimated values might be unstable, in which case using the maximum estimated value can lead to unreasonable bounds for $\alpha_{1}$. An easier to implement conservative alternative is to use the average derivatives with respect to $D$, $E\left[\frac{\partial E(Y|D=d,X=x,Y>0)}{\partial d}\right]$ and rely on bootstrapped standard errors for inference.
\end{remark}
 
\section{Empirical Illustration: \cite{lee1995semi}}\label{S7b}
In this section, we implement the proposed test to the data from \cite{lee1995semi}.\footnote{Data availability statement: The data that support the findings of this study are openly available in the Journal of Applied Econometrics Data Archive at http://dx.doi.org/10.15456/jae.2022313.1130270920.} Using the 1987 cross-section of the Michigan Panel Study of Income Dynamics, the authors study the responses of married women's labor supply ($Y$) - measured in hours per year - to hundreds of dollars in ``other'' household income ($D$), which is endogenous. The instrumental variables explored are dummy variables for the husband's occupation ($Z$), which implies the restrictive identifying assumption that the wife's labor supply is affected by the husband's occupation only through their income. Following the original study, we add other covariates ($X$) in the linear index for both the outcome and treatment equations, controlling for factors that could impact women's labor supply. Those include a quadratic on the person's age, their years of completed education, the number of children coded in three categories (children up to 5 years old, ages 6 to 13, ages 14 to 17), the local unemployment rate in percentage points, and indicators for race (0 if white, 1 otherwise), homeownership (1 if owner, 0 otherwise) and if the couple has a mortgage on their home (1 if yes, 0 otherwise).

Our empirical illustration considers as an instrument the binary variable indicating if the husband's occupation is classified as manager or professional.\footnote{The 1987 PSID uses 3-digit occupation codes from the 1970 U.S. Census, and this dummy variable seems to include workers listed in the categories ``1-195 Professional, Technical, and Kindred Workers,'' and ``201–245 Managers and Administrators, except Farm.''} 
 
Following \cite{lee1995semi}, we proceed with the analysis focusing on the data for married couples with non-negative family total income or ``other'' household income and where the wife was of working age (18-64) and not self-employed. These selections results in 3,277 observations, for which 26 percent of wage observations are censored. Table \ref{Table:Lee} presents the estimates obtained using the IV Tobit model.
 
\begin{table}[!htbp]
\caption{IV Tobit specification for \cite{lee1995semi}}
\begin{tabular}{l|cc} \label{Table:Lee}
& \multicolumn{2}{c}{MLE}   \\ 
 & Other household income & Hours per year worked\\
  \hline\hline\\
Husband's occupation: manager or professional  & $ 120.802^{***}$ & \\ 
                       & $(10.813)$ & \\
Other household income & & $-0.973^{***}$ \\
                             & & $(0.373)$\\
Age   & $13.686^{***}$ & $72.414^{***}$ \\
      & $(3.251)$   & $(15.249)$   \\
Age squared  & $-0.105^{***}$ & $ -1.221^{***}$ \\
     & $(0.039)$   & $(0.178)$   \\
Education  & $20.281^{***}$ & $92.107^{***}$ \\
     & $(2.071)$   & $(13.231)$   \\
Children under 5  & $9.448$ & $ -500.332^{***}$ \\
     & $(6.482)$   & $(28.163)$   \\
Children between 6 and 13  & $3.204$ & $ -211.687^{***}$ \\
     & $(5.601)$   & $(23.737)$   \\
Children between 14 and 17 & $12.881$ & $-16.878$ \\
     & $(9.342)$   & $(39.152)$   \\
Nonwhite  & $-59.550^{***}$ & $ 146.336^{***}$ \\
     & $(10.210)$   & $(51.105)$   \\
Homeowner  & $60.591^{***}$ & $13.830$ \\
     & $(15.461)$   & $(69.142)$   \\
Has mortgage  & $24.192^{*}$ & $ 254.660^{***}$ \\
     & $(13.954)$   & $(61.036)$   \\
Local Unemployment  & $-8.165^{***}$ & $ -41.979^{***}$ \\
     & $(1.941)$   & $(8.966)$   \\
Constant   & $-339.059^{***}$ & $-337.347$ \\
            & $(63.387)$   & $(314.207)$   \\
$\rho$ & &$0.042$ \\ 
             & &$(0.094)$ \\  
             $n$ &$3,377$ &$3,377$ \\ \hline \hline
\end{tabular}
\begin{center}
\scriptsize{Standard errors (in parentheses); ***: significant at 1\% level; *significant at 10\% level}
\end{center}
\label{application1}
\end{table}
The first column presents the first-stage estimates indicating the relevance of the potential $IV$. The second column presents the structural equation reflecting the effect the treatment variable has on the outcome equation. The parameter $\rho$ shows evidence of no correlation between the unobservables driving the ``other'' household income and hours worked after controlling for covariates. The estimated coefficient of interest indicates that other household income negatively affects the wife's labor supply after conditional on the household characteristics. In particular, for women working positive hours, an increase of one thousand dollars in household income from other sources is estimated to reduce hours worked by 9.7 hours per year. The direction of the impacts at the intensive margin of hours worked follows intuitive patterns and is qualitatively similar to those in \cite{lee1995semi}. 
However, the model is rejected at conventional significance levels when we test for the IV-Tobit model's validity,  ($\hat{\theta}_{0.99}= 0.1335>0$, $\hat{\theta}_{0.95}= 0.1368>0$, and $\hat{\theta}_{0.90}=0.1385>0$). This indicates that the assumptions underlying the IV-Tobit model are not compatible with the empirical distribution of the data, and caution is needed when relying on the results.

\subsection*{Bounds under Assumption \ref{R2} and latent linear index} 

 Given the rejection of the IV-Tobit model in this case, we relax the distributional and exogeneity of treatment assumptions, and construct lower bounds on the treatment effect under the MTS assumption and latent linear index only.  Assumption \ref{R2} imposes that average unobservables affecting women's preferences related to hours worked away from home, are monotonically decreasing in characteristics leading to higher income of other sources in the household, such as husband's income. In other words, households that prefer flexible schedules for women might similarly prioritize partner jobs that provide higher income.

 \begin{table}[!htbp]
\caption{Confidence sets for parameter of interest}
  \begin{tabular}{l|ccc} \label{Table:app2bounds}
 Parameter &  & IV-Tobit estimates & $\alpha_1$'s Lower Bound \\\hline\hline
$\alpha_1$ &  & -0.973  & -0.419 \\
           &  & (0.373) & (0.125)\\
\hline\hline
\end{tabular}
\begin{center}
\scriptsize{Note: Standard errors in parentheses.}
\end{center}
\end{table}
 The first column of Table \ref{Table:app2bounds} repeats the estimate for $\alpha_{1}$ from Table \ref{application1}. The second column reports the estimated lower bound for $\alpha_{1}$, obtained under Assumption \ref{R2} and latent linear index only based on the nonparametric estimate of the average derivative of the conditional expectation for yearly hours worked with respect to the household income from other sources. As described in Remark \ref{remcov}, we opt to use the estimate for the average derivatives due to instability of the estimated derivatives across the different values of covariates and treatment. This is a conservative approach regarding the bounds for $\alpha$. Inference for the average derivative can be obtained by bootstrapping. Even after relaxing the normality of errors and treatment exogeneity, the lower bound for the size of the effect of having higher household income from sources other than the wife's labor on their labor supply indicates an effect larger than -4.19 hours worked per year for married women. Hence, we can rule out annual reductions of more than 4.2 hours in female labor supply for every one thousand dollars in other household income, but cannot reject that the effect is zero or positive.

\section{Conclusion}\label{S8}

In this paper, we develop sharp testable equalities for the classic Tobit and IV-Tobit models that can detect all observable violations of the model's assumptions. The results are shown to extend to many other popular Tobit-type ``two-part'' models. By converting these sharp equalities into conditional moment inequalities, we propose a testing procedure that detects violations of the Tobit model assumptions on a grid on the joint support of the outcome (and treatment) variables, leveraging inference results from \cite{chernozhukov_intersection_2013} and the implementation from \cite{chernozhukov_implementing_2015}.
\par 
Simulation results suggest the test performs well for reasonably sized samples (larger than 5000 observations). The test is conservative for smaller samples, over-rejecting the null hypothesis of model validity. Simulations indicate that the test is powerful to detect violations of the exogeneity assumption for the treatment/instrument that affect the point estimates and inference. Finally, the proposed test exhibits good performance for violations of the distributional assumptions about the error structure.
\par 
We provide a brief review of existing models that could be implemented under weak/alternative assumptions when the Tobit model is rejected. Furthermore, we propose a simple model that partially identifies the parameter of interest by relying solely on linear index and monotone treatment selection restrictions, a standard assumption from the partial identification literature \citep{manski2000amonotone}.
\par
We illustrate our methods on data from \cite{lee1995semi}. We replicate qualitatively the results in the original paper and the proposed test for validity of the IV-Tobit model rejects the null hypothesis in this empirical application. We estimate our proposed lower bound for the effect of household income from sources other than the wife's labor on their labor supply, which does not rely on the normality of latent errors or treatment exogeneity.  While we can rule out that an extra 1,000 dollars in other household income reduce female labor supply by more than 4.2 hours per year, we cannot rule out that the effect is zero.

\bibliographystyle{jpe}
\bibliography{references}

@article{chesher2023iv,
  title={IV methods for Tobit models},
  author={Chesher, Andrew and Kim, Dongwoo and Rosen, Adam M},
  journal={Journal of Econometrics},
  volume={235},
  number={2},
  pages={1700--1724},
  year={2023},
  publisher={Elsevier}
}

@misc{bai2024sharptestableimplicationsencouragement,
      title={Sharp Testable Implications of Encouragement Designs}, 
      author={Yuehao Bai and Max Tabord-Meehan},
      year={2024},
      eprint={2411.09808},
      archivePrefix={arXiv},
      primaryClass={econ.EM},
      url={https://arxiv.org/abs/2411.09808}, 
}

@article{goff2024testing,
  title={Testing Identifying Assumptions in Parametric Separable Models: A Conditional Moment Inequality Approach},
  author={Goff, Leonard and K{\'e}dagni, D{\'e}sir{\'e} and Wu, Huan},
  journal={arXiv preprint arXiv:2410.12098},
  year={2024}
}

@article{lee1995semi,
  title={Semi-parametric estimation of simultaneous equations with limited dependent variables: a case study of female labour supply},
  author={Lee, Myoung-Jae},
  journal={Journal of Applied Econometrics},
  volume={10},
  number={2},
  pages={187--200},
  year={1995},
  publisher={Wiley Online Library}
}

@article{kedagni2020generalized,
  title={Generalized instrumental inequalities: testing the instrumental variable independence assumption},
  author={K{\'e}dagni, D{\'e}sir{\'e} and Mourifi{\'e}, Ismael},
  journal={Biometrika},
  volume={107},
  number={3},
  pages={661--675},
  year={2020},
  publisher={Oxford University Press}
}

@article{okumura2014concave,
  title={Concave-monotone treatment response and monotone treatment selection: With an application to the returns to schooling},
  author={Okumura, Tsunao and Usui, Emiko},
  journal={Quantitative Economics},
  volume={5},
  number={1},
  pages={175--194},
  year={2014},
  publisher={Wiley Online Library}
}

@article{manski1997monotone,
  title={Monotone treatment response},
  author={Manski, Charles F},
  journal={Econometrica: Journal of the Econometric Society},
  pages={1311--1334},
  year={1997},
  publisher={JSTOR}
}

@article{jiang2014monotone,
  title={Monotone confounding, monotone treatment selection and monotone treatment response},
  author={Jiang, Zhichao and Chiba, Yasutaka and VanderWeele, Tyler J},
  journal={Journal of causal inference},
  volume={2},
  number={1},
  pages={1--12},
  year={2014},
  publisher={De Gruyter}
}

@article{honore1993orthogonality,
  title={Orthogonality conditions for Tobit models with fixed effects and lagged dependent variables},
  author={Honor{\'e}, Bo E},
  journal={Journal of econometrics},
  volume={59},
  number={1-2},
  pages={35--61},
  year={1993},
  publisher={Elsevier}
}

@article{amemiya1979estimation,
  title={The estimation of a simultaneous-equation Tobit model},
  author={Amemiya, Takeshi},
  journal={International economic review},
  pages={169--181},
  year={1979},
  publisher={JSTOR}
}

@techreport{heckman1977dummy,
  title={Dummy endogenous variables in a simultaneous equation system},
  author={Heckman, James J},
  year={1977},
  institution={National Bureau of Economic Research}
}

@article{heckman1979sample,
  title={Sample selection bias as a specification error},
  author={Heckman, James J},
  journal={Econometrica: Journal of the econometric society},
  pages={153--161},
  year={1979},
  publisher={JSTOR}
}

@article{manski2000amonotone,
author = {Manski, Charles F. and Pepper, John V.},
title = {Monotone Instrumental Variables: With an Application to the Returns to Schooling},
journal = {Econometrica},
volume = {68},
number = {4},
pages = {997-1010},
doi = {https://doi.org/10.1111/1468-0262.t01-1-00144a},
url = {https://onlinelibrary.wiley.com/doi/abs/10.1111/1468-0262.t01-1-00144a},
eprint = {https://onlinelibrary.wiley.com/doi/pdf/10.1111/1468-0262.t01-1-00144a},
year = {2000}
}

@article{honore2000estimation,
  title={Estimation of Tobit-type models with individual specific effects},
  author={Honore, Bo E and Kyriazidou, Ekaterini and Powell, JL},
  journal={Econometric reviews},
  volume={19},
  number={3},
  pages={341--366},
  year={2000},
  publisher={Taylor \& Francis}
}

@article{wooldridge2005simple,
  title={Simple solutions to the initial conditions problem in dynamic, nonlinear panel data models with unobserved heterogeneity},
  author={Wooldridge, Jeffrey M},
  journal={Journal of applied econometrics},
  volume={20},
  number={1},
  pages={39--54},
  year={2005},
  publisher={Wiley Online Library}
}

@article{carson2007tobit,
  title={The Tobit model with a non-zero threshold},
  author={Carson, Richard T and Sun, Yixiao},
  journal={The Econometrics Journal},
  volume={10},
  number={3},
  pages={488--502},
  year={2007},
  publisher={Oxford University Press Oxford, UK}
}

@article{barros2018generalized,
  title={Generalized Tobit models: Diagnostics and application in econometrics},
  author={Barros, Michelli and Galea, Manuel and Leiva, V{\'\i}ctor and Santos-Neto, Manoel},
  journal={Journal of Applied Statistics},
  volume={45},
  number={1},
  pages={145--167},
  year={2018},
  publisher={Taylor \& Francis}
}

@article{chesher2017generalized,
  title={Generalized instrumental variable models},
  author={Chesher, Andrew and Rosen, Adam M},
  journal={Econometrica},
  volume={85},
  number={3},
  pages={959--989},
  year={2017},
  publisher={Wiley Online Library}
}

@article{das2002estimators,
  title={Estimators and inference in a censored regression model with endogenous covariates},
  author={Das, Mitali},
  year={2002}
}

@article{chernozhukov2015quantile,
  title={Quantile regression with censoring and endogeneity},
  author={Chernozhukov, Victor and Fern{\'a}ndez-Val, Iv{\'a}n and Kowalski, Amanda E},
  journal={Journal of Econometrics},
  volume={186},
  number={1},
  pages={201--221},
  year={2015},
  publisher={Elsevier}
}

@article{blundell2007censored,
  title={Censored regression quantiles with endogenous regressors},
  author={Blundell, Richard and Powell, James L},
  journal={Journal of Econometrics},
  volume={141},
  number={1},
  pages={65--83},
  year={2007},
  publisher={Elsevier}
}

@article{honore1994pairwise,
  title={Pairwise difference estimators of censored and truncated regression models},
  author={Honor{\'e}, Bo E and Powell, James L},
  journal={Journal of Econometrics},
  volume={64},
  number={1-2},
  pages={241--278},
  year={1994},
  publisher={Elsevier}
}

@article{honore2020selection,
  title={Selection without exclusion},
  author={Honor{\'e}, Bo E and Hu, Luojia},
  journal={Econometrica},
  volume={88},
  number={3},
  pages={1007--1029},
  year={2020},
  publisher={Wiley Online Library}
}

@article{powell1986symmetrically,
  title={Symmetrically trimmed least squares estimation for Tobit models},
  author={Powell, James L},
  journal={Econometrica: journal of the Econometric Society},
  pages={1435--1460},
  year={1986},
  publisher={JSTOR}
}

@article{cragg1971some,
  title={Some statistical models for limited dependent variables with application to the demand for durable goods},
  author={Cragg, John G},
  journal={Econometrica: journal of the Econometric Society},
  pages={829--844},
  year={1971},
  publisher={JSTOR}
}

@article{POWELL1984303,
title = {Least absolute deviations estimation for the censored regression model},
journal = {Journal of Econometrics},
volume = {25},
number = {3},
pages = {303-325},
year = {1984},
issn = {0304-4076},
doi = {https://doi.org/10.1016/0304-4076(84)90004-6},
url = {https://www.sciencedirect.com/science/article/pii/0304407684900046},
author = {James L Powell}
}

@article{newey1987efficient,
  title={Efficient estimation of limited dependent variable models with endogenous explanatory variables},
  author={Newey, Whitney K},
  journal={Journal of econometrics},
  volume={36},
  number={3},
  pages={231--250},
  year={1987},
  publisher={Elsevier}
}

@article{arai_testing_2022,
	title = {Testing identifying assumptions in fuzzy regression discontinuity designs},
	volume = {13},
	issn = {1759-7323},
	url = {http://qeconomics.org/ojs/index.php/qe/article/view/QE1367},
	doi = {10.3982/QE1367},
	language = {en},
	number = {1},
	urldate = {2024-02-16},
	journal = {Quantitative Economics},
	author = {Arai, Yoichi and Hsu, Yu-Chin and Kitagawa, Toru and Mourifié, Ismael and Wan, Yuanyuan},
	year = {2022},
	pages = {1--28},
}

@article{calonico2018effect,
  title={On the effect of bias estimation on coverage accuracy in nonparametric inference},
  author={Calonico, Sebastian and Cattaneo, Matias D and Farrell, Max H},
  journal={Journal of the American Statistical Association},
  volume={113},
  number={522},
  pages={767--779},
  year={2018},
  publisher={Taylor \& Francis}
}

@article{calonico2019nprobust,
  title={nprobust: Nonparametric Kernel-Based Estimation and Robust Bias-Corrected Inference},
  author={Calonico, Sebastian and Cattaneo, Matias D and Farrell, Max H},
  journal={Journal of statistical software},
  volume={91},
  number={8},
  pages={1--33},
  year={2019}
}

@article{calonico2022coverage,
  title={Coverage error optimal confidence intervals for local polynomial regression},
  author={Calonico, Sebastian and Cattaneo, Matias D and Farrell, Max H},
  journal={Bernoulli},
  volume={28},
  number={4},
  year={2022}
}

@article{chernozhukov_implementing_2015,
	title = {Implementing {Intersection} {Bounds} in {Stata}},
	volume = {15},
	issn = {1536-867X, 1536-8734},
	url = {http://journals.sagepub.com/doi/10.1177/1536867X1501500103},
	doi = {10.1177/1536867X1501500103},
	language = {en},
	number = {1},
	urldate = {2024-02-16},
	journal = {The Stata Journal: Promoting communications on statistics and Stata},
	author = {Chernozhukov, Victor and Kim, Wooyoung and Lee, Sokbae and Rosen, Adam M.},
	month = apr,
	year = {2015},
	pages = {21--44},
}

@article{chernozhukov_intersection_2013,
	title = {Intersection {Bounds}: {Estimation} and {Inference}},
	volume = {81},
	issn = {0012-9682},
	shorttitle = {Intersection {Bounds}},
	url = {http://doi.wiley.com/10.3982/ECTA8718},
	doi = {10.3982/ECTA8718},
	language = {en},
	number = {2},
	urldate = {2024-02-15},
	journal = {Econometrica},
	author = {Chernozhukov, Victor and Lee, Sokbae and Rosen, Adam M.},
	year = {2013},
	pages = {667--737},
}

@article{gunsilius_nontestability_2021,
	title = {Nontestability of instrument validity under continuous treatments},
	volume = {108},
	issn = {0006-3444, 1464-3510},
	url = {https://academic.oup.com/biomet/article/108/4/989/6035117},
	doi = {10.1093/biomet/asaa101},
	language = {en},
	number = {4},
	urldate = {2024-02-15},
	journal = {Biometrika},
	author = {Gunsilius, F F},
	month = nov,
	year = {2021},
	pages = {989--995},
}

@article{huber_testing_2015,
	title = {Testing {Instrument} {Validity} for {LATE} {Identification} {Based} on {Inequality} {Moment} {Constraints}},
	volume = {97},
	issn = {0034-6535, 1530-9142},
	url = {https://direct.mit.edu/rest/article/97/2/398-411/58220},
	doi = {10.1162/REST_a_00450},
	language = {en},
	number = {2},
	urldate = {2024-02-15},
	journal = {Review of Economics and Statistics},
	author = {Huber, Martin and Mellace, Giovanni},
	month = may,
	year = {2015},
	pages = {398--411},
}

@article{kitagawa_test_2015,
	title = {A {Test} for {Instrument} {Validity}},
	volume = {83},
	issn = {0012-9682},
	url = {https://www.econometricsociety.org/doi/10.3982/ECTA11974},
	doi = {10.3982/ECTA11974},
	language = {en},
	number = {5},
	urldate = {2024-02-15},
	journal = {Econometrica},
	author = {Kitagawa, Toru},
	year = {2015},
	pages = {2043--2063},
}

@article{kedagni_generalized_2020,
	title = {Generalized instrumental inequalities: testing the instrumental variable independence assumption},
	volume = {107},
	issn = {0006-3444, 1464-3510},
	shorttitle = {Generalized instrumental inequalities},
	url = {https://academic.oup.com/biomet/article/107/3/661/5767137},
	doi = {10.1093/biomet/asaa003},
	language = {en},
	number = {3},
	urldate = {2024-02-15},
	journal = {Biometrika},
	author = {Kédagni, Désiré and Mourifié, Ismael},
	month = sep,
	year = {2020},
	pages = {661--675},
}

@article{mourifie_testing_2017,
	title = {Testing {Local} {Average} {Treatment} {Effect} {Assumptions}},
	volume = {99},
	issn = {0034-6535, 1530-9142},
	url = {https://direct.mit.edu/rest/article/99/2/305-313/58389},
	doi = {10.1162/REST_a_00622},
	language = {en},
	number = {2},
	urldate = {2024-02-15},
	journal = {The Review of Economics and Statistics},
	author = {Mourifié, Ismael and Wan, Yuanyuan},
	month = may,
	year = {2017},
	pages = {305--313},
}

@book{wooldridge_econometric_2010,
	address = {Cambridge, Mass},
	edition = {2nd ed},
	title = {Econometric analysis of cross section and panel data},
	isbn = {978-0-262-23258-6},
	publisher = {MIT Press},
	author = {Wooldridge, Jeffrey M.},
	year = {2010},
	note = {OCLC: ocn627701062},
}

@article{tobin_estimation_1958,
	title = {Estimation of {Relationships} for {Limited} {Dependent} {Variables}},
	volume = {26},
	issn = {00129682},
	url = {https://www.jstor.org/stable/1907382?origin=crossref},
	doi = {10.2307/1907382},
	number = {1},
	urldate = {2024-02-15},
	journal = {Econometrica},
	author = {Tobin, James},
	month = jan,
	year = {1958},
	pages = {24},
}

@article{tauchen_diagnostic_1985,
	title = {Diagnostic testing and evaluation of maximum likelihood models},
	volume = {30},
	issn = {03044076},
	url = {https://linkinghub.elsevier.com/retrieve/pii/0304407685901496},
	doi = {10.1016/0304-4076(85)90149-6},
	language = {en},
	number = {1-2},
	urldate = {2024-02-15},
	journal = {Journal of Econometrics},
	author = {Tauchen, George},
	month = oct,
	year = {1985},
	pages = {415--443},
}

@article{smith_exogeneity_1986,
	title = {An {Exogeneity} {Test} for a {Simultaneous} {Equation} {Tobit} {Model} with an {Application} to {Labor} {Supply}},
	volume = {54},
	issn = {00129682},
	url = {https://www.jstor.org/stable/1911314?origin=crossref},
	doi = {10.2307/1911314},
	number = {3},
	urldate = {2024-02-15},
	journal = {Econometrica},
	author = {Smith, Richard J. and Blundell, Richard W.},
	month = may,
	year = {1986},
	pages = {679},
}

@article{reynolds_testing_1991,
	title = {Testing and correcting for distributional misspecifications in the {Tobit} model: {An} application of the {Information} {Matrix} test},
	volume = {16},
	issn = {0377-7332, 1435-8921},
	shorttitle = {Testing and correcting for distributional misspecifications in the {Tobit} model},
	url = {http://link.springer.com/10.1007/BF01206278},
	doi = {10.1007/BF01206278},
	language = {en},
	number = {3},
	urldate = {2024-02-15},
	journal = {Empirical Economics},
	author = {Reynolds, A. and Shonkwiler, J. S.},
	month = sep,
	year = {1991},
	pages = {313--323},
}

@article{newey_maximum_1985,
	title = {Maximum {Likelihood} {Specification} {Testing} and {Conditional} {Moment} {Tests}},
	volume = {53},
	issn = {00129682},
	url = {https://www.jstor.org/stable/1911011?origin=crossref},
	doi = {10.2307/1911011},
	number = {5},
	urldate = {2024-02-15},
	journal = {Econometrica},
	author = {Newey, Whitney K.},
	month = sep,
	year = {1985},
	pages = {1047},
}

@article{newey_specification_1987,
	title = {Specification tests for distributional assumptions in the {Tobit} model},
	volume = {34},
	issn = {03044076},
	url = {https://linkinghub.elsevier.com/retrieve/pii/0304407687900704},
	doi = {10.1016/0304-4076(87)90070-4},
	language = {en},
	number = {1-2},
	urldate = {2024-02-15},
	journal = {Journal of Econometrics},
	author = {Newey, Whitney K.},
	month = jan,
	year = {1987},
	pages = {125--145},
}

@article{nelson1978specification,
  title={Specification and estimation of a simultaneous-equation model with limited dependent variables},
  author={Nelson, Forrest and Olson, Lawrence},
  journal={International Economic Review},
  pages={695--709},
  year={1978},
  publisher={JSTOR}
}

@article{nelson_test_1981,
	title = {A {Test} for {Misspecification} in the {Censored} {Normal} {Model}},
	volume = {49},
	issn = {00129682},
	url = {https://www.jstor.org/stable/1912756?origin=crossref},
	doi = {10.2307/1912756},
	number = {5},
	urldate = {2024-02-15},
	journal = {Econometrica},
	author = {Nelson, Forrest D.},
	month = sep,
	year = {1981},
	pages = {1317},
}

@article{lin_test_1984,
	title = {A {Test} of the {Tobit} {Specification} {Against} an {Alternative} {Suggested} by {Cragg}},
	volume = {66},
	issn = {00346535},
	url = {https://www.jstor.org/stable/1924712?origin=crossref},
	doi = {10.2307/1924712},
	number = {1},
	urldate = {2024-02-15},
	journal = {The Review of Economics and Statistics},
	author = {Lin, Tsai-Fen and Schmidt, Peter},
	month = feb,
	year = {1984},
	pages = {174},
}

@book{li_nonparametric_2023,
	address = {Princeton Oxford},
	edition = {First paperback printing},
	title = {Nonparametric econometrics: theory and practice},
	isbn = {978-0-691-24808-0},
	shorttitle = {Nonparametric econometrics},
	language = {eng},
	publisher = {Princeton University Press},
	author = {Li, Qi and Racine, Jeffrey},
	year = {2023},
}

@article{holden_testing_2004,
	title = {Testing the {Normality} {Assumption} in the {Tobit} {Model}},
	volume = {31},
	issn = {0266-4763, 1360-0532},
	url = {http://www.tandfonline.com/doi/abs/10.1080/02664760410001681783},
	doi = {10.1080/02664760410001681783},
	language = {en},
	number = {5},
	urldate = {2024-02-15},
	journal = {Journal of Applied Statistics},
	author = {Holden, Darryl},
	month = jun,
	year = {2004},
	pages = {521--532},
}

@article{drukker_bootstrapping_2002,
	title = {Bootstrapping a {Conditional} {Moments} {Test} for {Normality} after {Tobit} {Estimation}},
	volume = {2},
	issn = {1536-867X, 1536-8734},
	url = {http://journals.sagepub.com/doi/10.1177/1536867X0200200202},
	doi = {10.1177/1536867X0200200202},
	language = {en},
	number = {2},
	urldate = {2024-02-15},
	journal = {The Stata Journal: Promoting communications on statistics and Stata},
	author = {Drukker, David M.},
	month = jun,
	year = {2002},
	pages = {125--139},
}

@article{bera_testing_1984,
	title = {Testing the {Normality} {Assumption} in {Limited} {Dependent} {Variable} {Models}},
	volume = {25},
	issn = {00206598},
	url = {https://www.jstor.org/stable/2526219?origin=crossref},
	doi = {10.2307/2526219},
	number = {3},
	urldate = {2024-02-15},
	journal = {International Economic Review},
	author = {Bera, Anil K. and Jarque, Carlos M. and Lee, Lung-Fei},
	month = oct,
	year = {1984},
	pages = {563},
}

@article{acerenza_testing_2023,
	title = {Testing identifying assumptions in bivariate probit models},
	volume = {38},
	issn = {0883-7252, 1099-1255},
	url = {https://onlinelibrary.wiley.com/doi/10.1002/jae.2956},
	doi = {10.1002/jae.2956},
	language = {en},
	number = {3},
	urldate = {2024-02-15},
	journal = {Journal of Applied Econometrics},
	author = {Acerenza, Santiago and Bartalotti, Otávio and Kédagni, Désiré},
	month = apr,
	year = {2023},
	pages = {407--422},
}

@article{han_identification_2017,
	title = {Identification in a generalization of bivariate probit models with dummy endogenous regressors},
	volume = {199},
	issn = {03044076},
	url = {https://linkinghub.elsevier.com/retrieve/pii/S0304407617300465},
	doi = {10.1016/j.jeconom.2017.04.001},
	language = {en},
	number = {1},
	urldate = {2024-02-15},
	journal = {Journal of Econometrics},
	author = {Han, Sukjin and Vytlacil, Edward J.},
	month = jul,
	year = {2017},
	pages = {63--73},
}
\appendix
\newpage
\section{Sharpness for the classic Tobit case}\label{app0}
Let $\tilde{\alpha}_1, \tilde{\alpha}_0, \sigma^2$ be identified. Suppose (\ref{CTEQ2})-  (\ref{CTEQ1}) hold. Let $f_{\tilde{U}|D}= \frac{1}{\sqrt{\sigma^2 2\pi}}e^{-1/2 \frac{\tilde{U}^2}{\sigma^2}}$. Let $\tilde{Y}^*=\tilde{\alpha}_0+\tilde{\alpha}_1 D+\tilde{U}$ and $\tilde{Y}=\max(0,\tilde{Y}^*)$. 
From $f_{\tilde{U}|D}$ we can see assumption \ref{CT2} holds, since $f_{\tilde{U}|D}=f_{\tilde{U}}$. Also note $f_{\tilde{U}}$ is the $N(0,\sigma^2)$ density, thus assumption \ref{CT3} holds. 
Considering the positive values for $\tilde{Y}$, for any constants $c_{0},c_{1}$ such that $0<c_{0}<c_{1}$, then, 
\begin{eqnarray*}
P(c_0 \leq \tilde{Y}\leq c_1|D=d)&=&P(c_0 \leq \tilde{Y}^{*}\leq c_1|D=d)\\
&=&P(c_0 \leq \tilde{\alpha}_0+\tilde{\alpha}_1 d+\tilde{U}\leq c_1|D=d)\\
&=&P(c_0-\tilde{\alpha}_0-\tilde{\alpha}_1 d \leq \tilde{U} \leq c_1-\tilde{\alpha}_0-\tilde{\alpha}_1 d|D=d)\\
&=&P\left(\frac{c_0}{\sigma}-\frac{\tilde{\alpha}_0}{\sigma}-\frac{\tilde{\alpha}_1}{\sigma} d \leq \frac{\tilde{U}}{\sigma} \leq \frac{c_1}{\sigma}-\frac{\tilde{\alpha}_0}{\sigma}-\frac{\tilde{\alpha}_1}{\sigma} d\right)\\
&=&\Phi\left(\frac{c_1}{\sigma}-\frac{\tilde{\alpha}_0}{\sigma}-\frac{\tilde{\alpha}_1}{\sigma} d\right)-\Phi\left(\frac{c_0}{\sigma}-\frac{\tilde{\alpha}_0}{\sigma}-\frac{\tilde{\alpha}_1}{\sigma} d\right)\\
&=&P(c_0 \leq Y\leq c_1|D=d).
\end{eqnarray*}
The first three equalities follow from the definitions of $\tilde{Y}^*$ and $\tilde{Y}$. The fourth and fifth steps use the specific choice for the probability density of $\tilde{U}$, which implies independence from $D$ and normality respectively. The last step uses the relationship between the observed data and the model, established in Equation (\ref{CTEQ2}).
\par 
Furthermore, for $P(\tilde{Y}=0|D=d)$:
\begin{eqnarray*}
P(\tilde{Y}=0|D=d)&=&P(\tilde{Y}^{*}\leq 0|D=d)=P(\tilde{\alpha}_0+\tilde{\alpha}_1 d+\tilde{U}\leq 0|D=d)\\
&=&P(\tilde{U} \leq -\tilde{\alpha}_0-\tilde{\alpha}_1 d|D=d)=P(\tilde{U} \leq -\tilde{\alpha}_0-\tilde{\alpha}_1 d)\\
&=&\Phi\left(-\frac{\tilde{\alpha}_0}{\sigma}-\frac{\tilde{\alpha}_1}{\sigma} d\right)=1-\Phi\left(\frac{\tilde{\alpha}_0}{\sigma}+\frac{\tilde{\alpha}_1}{\sigma} d\right)=P(Y=0|D=d)
\end{eqnarray*}
The last step uses the equality established in Equation \ref{CTEQ1}. Thus, we characterized the distribution of $\tilde{Y}|D$, which is equal to the one of $Y|D$, with $D$ given. Thus, we can induce the observed distribution $Y,D$.
\section{Details of the IV-Tobit case}\label{app1}
\subsection{Derivation of equation (\ref{IVTEQ1})}
\begin{eqnarray}
&&P(c_0\leq Y \leq c_1, d_0 \leq D \leq d_1| Z=z)\nonumber\\
&=&P(c_0\leq Y \leq c_1, d_0 \leq D \leq d_1, Y^*\geq 0 | Z=z)+P(c_0\leq Y \leq c_1, d_0 \leq D \leq d_1, Y^*< 0 | Z=z) \nonumber \\
&=&P(c_0\leq Y^{*} \leq c_1, d_0 \leq D \leq d_1, Y^*\geq 0| Z=z)\nonumber \\
&=&P(c_0\leq \tilde{\beta}_0+\tilde{\beta}_1 z+\tilde{W} \leq c_1, d_0 \leq \tilde{\gamma}_0+\tilde{\gamma}_1 z+\tilde{V} \leq d_1, Y^*\geq 0| Z=z)\nonumber \\
&=& P(c_0-\tilde{\beta}_0-\tilde{\beta}_1 z\leq \tilde{W} \leq c_1-\tilde{\beta}_0-\tilde{\beta}_1 z, d_0-\tilde{\gamma}_0-\tilde{\gamma}_1 z \leq \tilde{V} \leq d_1-\tilde{\gamma}_0-\tilde{\gamma}_1 z, Y^*\geq 0 | Z=z) \nonumber \\ 
&=& P(c_0-\tilde{\beta}_0-\tilde{\beta}_1 z\leq \tilde{W} \leq c_1-\tilde{\beta}_0-\tilde{\beta}_1 z, d_0-\tilde{\gamma}_0-\tilde{\gamma}_1 z \leq \tilde{V} \leq d_1-\tilde{\gamma}_0-\tilde{\gamma}_1 z) \nonumber \\
&=& \Phi_{W,V}\left(\frac{c_1}{\sigma_{\tilde{W}}}-\frac{\tilde{\beta}_0}{\sigma_{\tilde{W}}}-\frac{\tilde{\beta}_1}{\sigma_{\tilde{W}}} z,\frac{d_1}{\sigma_{\tilde{V}}}-\frac{\tilde{\gamma}_0}{\sigma_{\tilde{V}}}-\frac{\tilde{\gamma}_1}{\sigma_{\tilde{V}}} z;\rho\right)-\Phi_{W,V}\left( \frac{c_0}{\sigma_{\tilde{W}}}-\frac{\tilde{\beta}_0}{\sigma_{\tilde{W}}}-\frac{\tilde{\beta}_1}{\sigma_{\tilde{W}}} z,\frac{d_1}{\sigma_{\tilde{V}}}-\frac{\tilde{\gamma}_0}{\sigma_{\tilde{V}}}-\frac{\tilde{\gamma}_1}{\sigma_{\tilde{V}}} z;\rho\right) \nonumber \\ 
&-& \Phi_{W,V}\left(\frac{c_1}{\sigma_{\tilde{W}}}-\frac{\tilde{\beta}_0}{\sigma_{\tilde{W}}}-\frac{\tilde{\beta}_1}{\sigma_{\tilde{W}}} z,\frac{d_0}{\sigma_{\tilde{V}}}-\frac{\tilde{\gamma}_0}{\sigma_{\tilde{V}}}-\frac{\tilde{\gamma}_1}{\sigma_{\tilde{V}}} z;\rho\right)+\Phi_{W,V}\left(\frac{c_0}{\sigma_{\tilde{W}}}-\frac{\tilde{\beta}_0}{\sigma_{\tilde{W}}}-\frac{\tilde{\beta}_1}{\sigma_{\tilde{W}}} z,\frac{d_0}{\sigma_{\tilde{V}}}-\frac{\tilde{\gamma}_0}{\sigma_{\tilde{V}}}-\frac{\tilde{\gamma}_1}{\sigma_{\tilde{V}}} z;\rho\right)\nonumber
\end{eqnarray}
The first equality follows from the law of total probability. The second through fourth equalities are in consequence of the model structure in \ref{IVT1}. The fifth step uses Assumption \ref{IVT2} and $c_0 \geq 0$. The final equality is by the properties of probabilities and the joint normality for $W, V$ (Assumption \ref{IVT3}).
\subsection{Derivation of equation (\ref{IVTEQ2})}
By a similar approach to the derivation of (\ref{IVTEQ1}):
\begin{eqnarray}
P(Y=0, d_0 \leq D \leq d_1| Z=z)&=&P(Y^* \leq 0, d_0 \leq D \leq d_1 | Z=z) \nonumber \\
&=& P(\tilde{W} \leq -\tilde{\beta}_0-\tilde{\beta}_1 z, d_0-\tilde{\gamma}_0-\tilde{\gamma}_1 \leq \tilde{V} \leq d_1-\tilde{\gamma}_0-\tilde{\gamma}_1 z) \nonumber \\ 
&=&  \Phi_{W,V}\left( -\frac{\tilde{\beta}_0}{\sigma_{\tilde{W}}}-\frac{\tilde{\beta}_1}{\sigma_{\tilde{W}}} z,\frac{d_1}{\sigma_{\tilde{V}}}-\frac{\tilde{\gamma}_0}{\sigma_{\tilde{V}}}-\frac{\tilde{\gamma}_1}{\sigma_{\tilde{V}}} z;\rho\right)\nonumber \\
&-&\Phi_{W,V}\left(-\frac{\tilde{\beta}_0}{\sigma_{\tilde{W}}}-\frac{\tilde{\beta}_1}{\sigma_{\tilde{W}}} z,\frac{d_0}{\sigma_{\tilde{V}}}-\frac{\tilde{\gamma}_0}{\sigma_{\tilde{V}}}-\frac{\tilde{\gamma}_1}{\sigma_{\tilde{V}}} z;\rho\right)
\end{eqnarray}
\subsection{Proof of Sharpness} 
Let $\beta_0,\beta_1, \gamma_0, \gamma_1, \rho$ be identified and equalities (\ref{IVTEQ1})-(\ref{IVTEQ3}) hold. Define the joint density of $\left(\tilde{W},\tilde{V}\right)$ conditional on $Z$ as
\begin{eqnarray*}
f_{\left(\tilde{W}, \tilde{V}\vert Z\right)}(w,v \vert z) = \frac{1}{\sqrt{1-\rho^2}}\phi\left(\frac{w-\rho v}{\sqrt{1-\rho^2}}\right)\phi(v),
\end{eqnarray*}
where $\phi(t) =\exp(-t^2/2)$, and define
\begin{eqnarray*}
\left\{ \begin{array}{lcl}
     \tilde{Y} &=& \max(0, \tilde{Y}^*) \\ \\
     \tilde{Y}^*&=& \beta_0+\beta_1 Z+\tilde{W} \\ \\ 
     \tilde{D} &=& \gamma_0+\gamma_1 Z+ \tilde{V}
     \end{array} \right.
\end{eqnarray*}
Note that $f_{\left(\tilde{W}, \tilde{V}\vert Z\right)}=f_{\left(\tilde{W}, \tilde{V}\right)}$, thus assumption \ref{IVT2} holds. Similarly, $\left(\tilde{W}, \tilde{V}\right)$ follow a bivariate normal distribution as $\begin{pmatrix}
\tilde{W}\\\tilde{V}
\end{pmatrix} \sim \mathcal N(\mu,\Sigma)$, where
$\mu=\begin{pmatrix}
0\\0
\end{pmatrix}$, and 
$\Sigma=
\begin{pmatrix}
1 & \rho  \\ 
 \rho & 1
\end{pmatrix}$. Let $\tilde{U}=\tilde{W}-\alpha_1\tilde{V}$, which implies $\tilde{U},\tilde{V}$ satisfies assumption \ref{IVT3}, given the scale-location normalizations.
\par 
Then, for any constants $0<c_0\leq c_1$ and $d_{0}<d_{1}$: 
\begin{eqnarray*}
&P&(c_0\leq \tilde{Y} \leq c_1, d_0 \leq \tilde{D} \leq d_1| Z=z)=P(c_0\leq \tilde{Y}^{*} \leq c_1, d_0 \leq \tilde{D} \leq d_1| Z=z)\\
&=&P(c_0\leq \beta_0+\beta_1 z+\tilde{W} \leq c_1, d_0 \leq \gamma_0+\gamma_1 z+\tilde{V} \leq d_1| Z=z)\\
&=&P(c_0-\beta_0-\beta_1 z\leq \tilde{W} \leq c_1-\beta_0-\beta_1 z, d_0-\gamma_0-\gamma_1 z \leq \tilde{V} \leq d_1-\gamma_0-\gamma_1 z| Z=z)\\
&=&P(c_0-\beta_0-\beta_1 z\leq \tilde{W} \leq c_1-\beta_0-\beta_1 z, d_0-\gamma_0-\gamma_1 z \leq \tilde{V} \leq d_1-\gamma_0-\gamma_1 z)\\
&=&\Phi_{W,V}( c_1-\beta_0-\beta_1 z,d_1-\gamma_0-\gamma_1 z;\rho)-\Phi_{W,V}( c_0-\beta_0-\beta_1 z,d_1-\gamma_0-\gamma_1 z;\rho) \\ 
&-& \Phi_{W,V}( c_1-\beta_0-\beta_1 z,d_0-\gamma_0-\gamma_1 z;\rho)+ \Phi_{W,V}( c_0-\beta_0-\beta_1 z,d_0-\gamma_0-\gamma_1 z;\rho)\\ 
&=& P(c_0\leq Y \leq c_1, d_0 \leq D \leq d_1| Z=z)
\end{eqnarray*}
The first through third equalities follow from the definitions of $\tilde{Y},\tilde{Y}^{*}$ and $\tilde{D}$. The fourth and fifth steps are consequences of the particular choice for joint density for $\left(\tilde{W},\tilde{V}\right)$ conditional on $Z$. The final equality is given by the relationship between observable data and the latent model structure in Equation (\ref{IVTEQ1}). Similar derivations hold for (\ref{IVTEQ1b})-(\ref{IVTEQ1c}). 

For the accumulation point, at $Y=0$:
\begin{eqnarray*}
&P&(\tilde{Y}=0, d_0 \leq \tilde{D} \leq d_1| Z=z)=P(\tilde{Y}^{*} \leq 0, d_0 \leq \tilde{D} \leq d_1| Z=z)\\
&=&P(\beta_0+\beta_1 z+\tilde{W} \leq 0, d_0 \leq \gamma_0+\gamma_1 z+\tilde{V} \leq d_1| Z=z)\\
&=&P(\tilde{W} \leq -\beta_0-\beta_1 z, d_0-\gamma_0-\gamma_1 z \leq \tilde{V} \leq d_1-\gamma_0-\gamma_1 z| Z=z)\\
&=&P(\tilde{W} \leq -\beta_0-\beta_1 z, d_0-\gamma_0-\gamma_1 z \leq \tilde{V} \leq d_1-\gamma_0-\gamma_1 z)\\
&=&\Phi_{W,V}(-\beta_0-\beta_1 z,d_1-\gamma_0-\gamma_1 z;\rho)-\Phi_{W,V}(-\beta_0-\beta_1 z,d_0-\gamma_0-\gamma_1 z;\rho)\\ 
&=& P(Y=0, d_0 \leq D \leq d_1| Z=z).
\end{eqnarray*}
The steps of the proof are similar to previous cases and the last equality is given by the relationship between observable data and the latent model structure in Equation \ref{IVTEQ2}. Similar derivations hold for equalities (\ref{IVTEQ2b})-(\ref{IVTEQ3}).

Thus, we characterized the distribution of $(\tilde{Y}, \tilde{D})| Z $, which coincides with the joint distribution of $(Y, D)|Z$, for given $Z$. Thus, we induced the observed distribution of the data $Y, D, Z$.

\section{Type 2 Tobit testable implications }\label{app4}
In this section, we derive the results of the Type 2 tobit model, also known as selection models or Heckman selection-type models \citep{heckman1979sample}. The basic setup with no covariates (which can be extended in several directions and with different distributional assumptions as well as to incorporate treatment endogeneity, as we pointed out for the classic Tobit model in proposition \ref{RemEx}) is:

 \begin{eqnarray}\label{SST0}
\left\{ \begin{array}{lcl}
     Y &=& Y^* \quad \text{if} \quad S=1 \\ \\
      Y &=& \text{missing} \quad \text{if} \quad S=0 \\ \\
     Y^* &=& \alpha_0+\alpha_1 D+U  \\ \\
     S&=& 1 \{ \gamma_0+\gamma_1 Z+V \geq 0 \}
     \end{array} \right.
\end{eqnarray}
Where $U, V$ are the latent structural error terms.

\begin{assumption}\label{SST2}
$D,Z$ be independent of $U, V$. Furthermore, let $\gamma_1 \neq 0$ 
\end{assumption}
\begin{assumption}\label{SST3}
Let $U, V$ follow a bivariate normal  distribution with covariance $\rho$, i.e., $\begin{pmatrix}
U\\V
\end{pmatrix} \sim \mathcal N(\mu,\Sigma)$, where
$\mu=\begin{pmatrix}
0\\0\
\end{pmatrix}$, and 
$\Sigma=
\begin{pmatrix}
\sigma^2_U & \rho_{UV}  \\ 
 \rho_{UV} & \sigma^2_V
\end{pmatrix}$.\end{assumption}

Note that $Y$ is missing at $S=0$. Thus, fully characterizing the distribution implies observing the behavior at the missing point and beyond it. 
\par 
From the observed data the conditional probabilities of $c_0 \leq Y\leq c_1$ for an $c_1,c_0$ can be computed. Then note for any $d \in \mathcal{D}$ and $z \in \mathcal{Z}$ :   
\begin{eqnarray}
 P(Y=\text{missing} \quad|D=d,Z=z)&=& P(S=0|D=d,Z=z) \nonumber \\
 &=& P( \gamma_0+\gamma_1 Z+V <0|D=d,Z=z) \nonumber \\
 &=& P( \gamma_0+\gamma_1 z+V <0) \nonumber \\
  &=& P(V < -\gamma_0-\gamma_1 z) \nonumber \\
 &=& \Phi_v\left(\frac{ -\gamma_0-\gamma_1 z}{\sigma_V}\right)
\end{eqnarray}
Where the first and the second equalities are due to the structure of the model described in equation (\ref{SST0}). The third step is due to Assumption \ref{SST2} and the last one follows from the normalization for normal random variables and Assumption \ref{SST3}.
\par 
Additionally, 
\begin{eqnarray}
P(c_0 \leq Y\leq c_1|D=d, Z=z)&=& P(c_0 \leq Y\leq c_1,S=0|D=d, Z=z) \nonumber \\
&+& P(c_0 \leq Y\leq c_1,S=1|D=d, Z=z) \nonumber \\
&=& P(c_0 \leq Y^*\leq c_1,S=1|D=d, Z=z) \nonumber \\
&=& P(c_0 \leq \alpha_0+\alpha_1 D+U \leq c_1,\gamma_0+\gamma_1 Z+V \geq 0|D=d, Z=z) \nonumber \\
&=& P(c_0 \leq \alpha_0+\alpha_1 d+U \leq c_1,\gamma_0+\gamma_1 z+V \geq 0|D=d, Z=z) \nonumber \\
&=& P(c_0-\alpha_0-\alpha_1 d \leq U \leq c_1- \alpha_0-\alpha_1 d,V \geq -\gamma_0-\gamma_1 z|D=d, Z=z) \nonumber \\
&=& P(c_0-\alpha_0-\alpha_1 d \leq U \leq c_1- \alpha_0-\alpha_1 d,V \geq -\gamma_0-\gamma_1 z) \nonumber \\
&=& \Phi_U\left( \frac{c_1- \alpha_0-\alpha_1 d}{\sigma_{U}}\right)- \Phi_U\left(\frac{c_0-\alpha_0-\alpha_1 d}{\sigma_{U}}\right) \nonumber \\
&-& \Phi_{U,V}\left(\frac{c_1- \alpha_0-\alpha_1 d}{\sigma_{U}},\frac{-\gamma_0-\gamma_1 z}{\sigma_{V}}, \rho_{UV}\right)  \nonumber \\
&+& \Phi_{U,V}\left(\frac{c_0- \alpha_0-\alpha_1 d}{\sigma_{U}},\frac{-\gamma_0-\gamma_1 z}{\sigma_{V}}, \rho_{UV}\right)
\end{eqnarray}
The first equality follows from the law of total probability. The second through fifth equalities follow from the model's structure described in Equation (\ref{SST0}), namely $P(Y>0, Y^*< 0|D=d)=0$ and $Y^{*}=Y, \text{not missing}$ for $S=1$, the structure of $S$, and the latent linear model. The sixth step is due to Assumption \ref{SST2}.
The last equality uses Assumption \ref{SST3} as well as properties of normal random variables. 
Thus, we can construct a test similarly to the one in the main text, using these equalities in addition to the ones of the form $P(Y\geq c_2|D=d, Z=z)$. 

\section{Results from Proposition \ref{RemEx}}\label{ApRemEx}
In this appendix, we derive the testable implications of the extensions to the classic Tobit and IV-Tobit discussed in Proposition \ref{RemEx}. We discuss testable implications assuming the identification of the relevant parameters or distributions.

\subsection{\cite{barros2018generalized} and \cite{carson2007Tobit} }
\cite{barros2018generalized} proposes a variant of the tobit model with elliptically contoured distributions and a non-zero threshold. At the same time \cite{carson2007Tobit} proposes a Tobit model with a non-zero threshold. In this section we combine both types of results and report the testable implications with a generic non-zero threshold and a generic known or identifiable parametric distribution function. In addition, extend the latent structure to be also known or an identifiable function up to a vector of parameters but invertible in $U$.  In this context, let: 

 \begin{eqnarray}\label{R1.1}
\left\{ \begin{array}{lcl}
     Y &=& \max(\tau,Y^*) \\ \\
     Y^* &=& g(\alpha,D,U)
     \end{array} \right.
\end{eqnarray}
\begin{assumption}\label{R1.2}
$D$ is independent of $U$.
\end{assumption}
\begin{assumption}\label{R1.3}
$U$ is distributed according to distribution $F_H(.)$ with parameters $\omega$.
\end{assumption}
In this context, the testable implications are, starting at the continuous part of the distribution of $Y$, the conditional probabilities of $c_0 \leq Y\leq c_1$ for a $c_1,c_0>\tau$ are observed. For any value of the treatment variables $d \in \mathcal{D}$:   
\begin{eqnarray}
P(c_0 \leq Y\leq c_1|D=d)&=&P(c_0 \leq Y\leq c_1,Y^* \geq \tau|D=d)+P(c_0 \leq Y\leq c_1,Y^*< \tau|D=d) \nonumber \\
&=& P(c_0 \leq Y\leq c_1,Y^*\geq \tau|D=d) \nonumber \\
&=& P(c_0 \leq Y^*\leq c_1,Y^* \geq \tau|D=d)  \nonumber \\
&=& P(c_0 \leq Y^*\leq c_1|D=d) \nonumber \\
&=&P(c_0 \leq g(\alpha,D,U) \leq c_1|D=d) \nonumber \\
&=& P(g^{-1}(\alpha,d,c_0) \leq U \leq g^{-1}(\alpha,d,c_1)|D=d) \nonumber \\
&=& F_H(g^{-1}(\alpha,d,c_1);\omega)- F_H(g^{-1}(\alpha,d,c_0);\omega)
\end{eqnarray}
Turning to the accumulation point, the observed event of $Y=0$,
\begin{eqnarray}
P(Y=0|D=d)=P(Y^* \leq \tau|D=d)=F_H(g^{-1}(\alpha,d,\tau); \omega).
\end{eqnarray}
These equalities can be used to construct a test in a similar way as did for the classic Tobit and IV-Tobit by adding the type of equalities $P(Y\geq c_2|D=d)$ which can be derived in a similar fashion. Also, note it is possible to extend the previous development to cases where the truncation takes the form $Y=Y^*$ if $ \tau_0 \leq Y^* \leq \tau_1$,  $Y=\tau_0$ if $ \tau_0 \geq Y^*$ and  $Y=\tau_1$ if $ Y^* \geq \tau_1$. 
\subsection{\cite{wooldridge2005simple, honore2000estimation, honore1993orthogonality}}
Consider the following dynamic version of the Tobit model, which is related to  \cite{wooldridge2005simple, honore2000estimation, honore1993orthogonality}. Here, we specify the conditional behaviour of $c_i$ in the spirit of \cite{wooldridge2005simple} and others. 

 \begin{eqnarray}\label{R1.1dyn}
\left\{ \begin{array}{lcl}
     Y_{i,t} &=& \max(0,Y_{i,t}^*) \\ \\
     Y_{i,t}^* &=& \alpha_0+\alpha_1D_{i,t}+\alpha_2g(Y_{i,t-1})+c_i+U_{i,t}
     \end{array} \right.
\end{eqnarray}
\begin{assumption}\label{R1.2dyn}
$D_{i,t}$ is independent of $U_{i,t}$ given $Y_{i,t-1},c_i$.
\end{assumption}
\begin{assumption}\label{R1.3dyn}
$U_{i,t}|Y_{i,t-1},c_i$ follows a $N(0,1)$ distribution. 
\end{assumption}
\begin{assumption}\label{R1.4dyn}
$c_i|Y_{i,t-1},D_{i,t-1}$ has a known distribution such as $N(0,1)$. 
\end{assumption}
Then, starting at the continuous part of the distribution of $Y$, the conditional probabilities of $c_0 \leq Y\leq c_1$ for a $c_1,c_0>0$ are observed. For any value of the treatment variables $d \in \mathcal{D}$, $P(c_0 \leq Y_{i,t}\leq c_1|D_{i,t}=d, Y_{i,t-1}=y)$:   
\begin{eqnarray}
&=& \int P(c_0 \leq Y_{i,t}\leq c_1|D_{i,t}=d, Y_{i,t-1}=y,c_i)f(c_i|D_{i,t}=d, Y_{i,t-1}=y)dc_i \nonumber \\
&=&  \int P(c_0 \leq Y_{i,t}\leq c_1|Y_{i,t-1}=y,c_i)f(c_i|D_{i,t}=d, Y_{i,t-1}=y)dc_i \nonumber \\
&=&  \int P(c_0 \leq Y^*_{i,t}\leq c_1|Y_{i,t-1}=y,c_i)f(c_i|D_{i,t}=d, Y_{i,t-1}=y)dc_i \nonumber \\
&=&  \int P(c_0-\alpha_0-\alpha_1d-\alpha_2g(y)-c_i \leq U_{i,t}\leq c_1-\alpha_0-\alpha_1d-\alpha_2g(y)-c_i|Y_{i,t-1}=y,c_i)f(c_i|d,y)dc_i \nonumber \\
&=&  \int[\Phi(c_1-\alpha_0-\alpha_1d-\alpha_2g(y)-c_i)-\Phi(c_0-\alpha_0-\alpha_1d-\alpha_2g(y)-c_i)]  \phi(c_i)dc_i 
\end{eqnarray}
Similarly, 

\begin{eqnarray}
P(Y_{i,t}=0|D_{i,t}=d, Y_{i,t-1}=y)&=& \int P(Y_{i,t}=0|D_{i,t}=d, Y_{i,t-1}=y,c_i)f(c_i|D_{i,t}=d, Y_{i,t-1}=y)dc_i \nonumber \\
&=&  \int P(Y_{i,t}=0|Y_{i,t-1}=y,c_i)f(c_i|D_{i,t}=d, Y_{i,t-1}=y)dc_i \nonumber \\
&=&  \int P(Y^*_{i,t}\leq 0|Y_{i,t-1}=y,c_i)f(c_i|D_{i,t}=d, Y_{i,t-1}=y)dc_i \nonumber \\
&=&  \int P(U_{i,t}\leq -\alpha_0-\alpha_1d-\alpha_2g(y)-c_i|Y_{i,t-1}=y,c_i)f(c_i|d,y)dc_i \nonumber \\
&=&  \int[\Phi(-\alpha_0-\alpha_1d-\alpha_2g(y)-c_i)]  \phi(c_i)dc_i 
\end{eqnarray}
Which can then be used to construct a test in a similar way as presented in the main text by adding the type of equalities $P(Y_{i,t}\geq c_2|D_{i,t}=d, Y_{i,t-1}=y)$, with the caveat that the right-hand side requires numerical integration or an approximation by an estimator when no closed form is available. Such estimator should ensure that the left-hand side of the equality converges at root-$N$ in order for the estimation step of the null model to be asymptotically negligible \cite[Appendix B]{acerenza_testing_2023}.

\section{Non-learnability}\label{ApLearn}
\subsection{Non-Learnability for the classic Tobit model}
As mentioned in Remark \ref{Rem:NLClassic}, the standard Tobit model is non-verifiable, that is, we cannot learn if the maintained Tobit model is the true data generating process based on the sharp equalities proposed in Section \ref{S91}. One can show that by finding an alternative model that is compatible with the equalities (\ref{CTEQ2})-(\ref{CTEQ1}) in all cases in which the Tobit model could not be disregarded.
Thus the model is non-verifiable in the sense that we can always construct a joint probability law of  $(\tilde{Y}, D,\tilde{U}$) that violates the Tobit model validity but satisfies the equalities. Concretely, for the classic Tobit model suppose that,  
\begin{eqnarray}
\left\{ \begin{array}{lcl}
     \bar{Y} &=& \max(0,\bar{Y}^*) \\ \\
     \bar{Y}^* &=& \alpha_0+\alpha_1 D+\tilde{U}
     \end{array} \right.
\end{eqnarray}
$\tilde{U}$ is a random variable that has the following mixed discrete-continuous ``density'':
\begin{eqnarray*}
 f_{\tilde{U}|D}(\tilde{u}|d)= \begin{cases}
\Phi(\hat{u})(\frac{e^d}{1+e^d}) \text{ if }  \tilde{u}=\hat{u} \\
 \Phi(-\alpha_0-\alpha_1d)-\Phi(\hat{u})(\frac{e^d}{1+e^d}) \text{ if }  \tilde{u} = -\alpha_0-\alpha_1d  \\
\phi(\tilde{u}) \text{ if } \tilde{u} \geq -\alpha_0-\alpha_1d\\
 0,\text{ otherwise}
 \end{cases}   
\end{eqnarray*}
Where $\hat{u}$ is some real number strictly less than $ -\alpha_0-\alpha_1d$, $\phi(\tilde{u})$ is the standard normal pdf and $\Phi(\tilde{u})$ is the standard normal cdf. Note that this is a valid distribution since (i) it integrates to $1$ from minus infinity to plus infinity and (ii)  $\Phi(\hat{u})(\frac{e^d}{1+e^d})>0,\phi(\tilde{u})>0$ and $\Phi(-\alpha_0-\alpha_1d)-\Phi(\hat{u})(\frac{e^d}{1+e^d}) >0$. This distribution assigns probability mass to two points for which the \emph{observed} outcome $\bar{Y}$ equals zero, $\hat{u}$ and $-\alpha_0-\alpha_1d$ while assigning density above $-\alpha_0-\alpha_1d$ by the usual standard normal density. Thus above $-\alpha_0-\alpha_1d$, $\tilde{U}$ behaves like a normal random variable, but below $-\alpha_0-\alpha_1d$ it behaves like a discrete random variable with two mass points.

For $c_0 \leq Y\leq c_1$ for a $c_1,c_0>0$ and for any value of the treatment variables $d \in \mathcal{D}$, assume that the sharp equalities hold:   
\begin{eqnarray*}
P(c_0 \leq Y\leq c_1|D=d)&=& \Phi(c_1-\alpha_0-\alpha_1 d)-\Phi(c_0-\alpha_0-\alpha_1 d)\\
P(Y=0|D=d)&=&1-\Phi(\alpha_0+\alpha_1 d).
\end{eqnarray*}
 Now define $\Bar{Y}_d=\alpha_0+\alpha_1 d+\tilde{U}$ 
\par 
In the above proposed DGP, the latent model is not normal. However, the DGP is compatible with the data and the testable implications derived in Section \ref{S9} since, 
\begin{eqnarray}
P(\bar{Y}_d< 0) &=&  P(\tilde{U}\leq -\alpha_0-\alpha_1d)\\
&=& \Phi(\hat{u})\left(\frac{e^d}{1+e^d}\right)+  \Phi(-\alpha_0-\alpha_1d)-\Phi(\hat{u})\left(\frac{e^d}{1+e^d}\right) \\
&=& \Phi(-\alpha_0-\alpha_1d)=1-\Phi(\alpha_0+\alpha_1d)\\
&=& P(Y=0|D=d)
\end{eqnarray}
The last equality is due to the testable equalities holding. Similarly for any $c_0,c_1>0$, 
\begin{eqnarray}
P(c_0 \leq \bar{Y}_d\leq c_1)&=& \Phi(c_{1}-\alpha_0-\alpha_1 d)-\Phi(c_0-\alpha_0-\alpha_1 d)\nonumber \\
&=&P(c_{0} \leq Y\leq c_{1}|D=d).
\end{eqnarray}
Hence, when the sharp equalities hold, there exists an alternative observationally equivalent model, to the classic tobit that can induce the observed data distribution. Indeed, the observed distribution of $Y$ and the non refuted marginal distribution of $U$, do not imply that we can learn the true distribution of ($Y_1,Y_0,U$).

\subsection{Non-Learnability for the IV- Tobit model}
As mentioned in proposition \ref{Rem:NLIV}, the IV-Tobit model is non-verifiable, that is, we cannot learn if the maintained  model is the true data generating process based on the sharp equalities proposed in Section \ref{S92}. One can show that by finding an alternative model that is compatible with the equalities (\ref{IVTEQ1})-(\ref{IVTEQ3}) in all cases in which the IV-Tobit model could not be disregarded.
Suppose that, 
\begin{eqnarray}\label{IVTNT1}
\left\{ \begin{array}{lcl}
     \tilde{Y} &=& \max(0,\tilde{Y}^*) \\ \\
     \tilde{Y}^* &=& \beta_0+\beta_1 Z+\tilde{W}  \\ \\
     \tilde{D}&=& \gamma_0+\gamma_1 Z+\tilde{V}
     \end{array} \right.
\end{eqnarray}
$\tilde{W}, \tilde{V}$ are  random variables that have the next mixed discrete-continuous behavior:
\begin{eqnarray*}
 f_{ \tilde{V}}(\tilde{v})= 
\phi(\tilde{v})  
\end{eqnarray*}
Where $\phi(\tilde{v})$ is the normal density. 
Furthermore, 
\begin{eqnarray*}
 f_{\tilde{W}| \tilde{V},Z}(\tilde{w}| \tilde{v},z)= \begin{cases}
\Phi(\hat{w}|\tilde{v};\rho)(\frac{e^z}{1+e^z}) \text{ if }  \tilde{w}=\hat{w}  \\
 \Phi(-\beta_0-\beta_1z|\tilde{v};\rho)-\Phi(\hat{w}|\tilde{v};\rho)(\frac{e^z}{1+e^z}) \text{ if }  \tilde{w} = -\beta_0-\beta_1z   \\
\phi(\tilde{w}| \tilde{v};\rho) \text{ if } \tilde{w} \geq -\beta_0-\beta_1z\\
 0,\text{ otherwise.}
 \end{cases}   
\end{eqnarray*}
Where $\hat{w}$ is some real number strictly less than $ -\beta_0-\beta_1z$, $\phi(\tilde{w}|\tilde{v},\rho)$ is the  normal p.d.f. conditional on $\tilde{v}$ with correlation coefficient $\rho$ and $\Phi(\tilde{w}|\tilde{v};\rho)$ is the  normal c.d.f. analog. Note that, $\Phi(\hat{w}|\tilde{v};\rho)(\frac{e^z}{1+e^z})>0,\phi(\tilde{w}|\tilde{v}; \rho)>0$ and $\Phi(-\beta_0-\beta_1z|\tilde{v};\rho)-\Phi(\hat{w}|\tilde{v}; \rho)(\frac{e^z}{1+e^z}) >0$. This distribution assigns probability mass to two points, $\hat{w}$ and $-\beta_0-\beta_1z$ while assigning the conditional normal density above $-\beta_0-\beta_1z$. Thus, above $-\beta_0-\beta_1z$, $\tilde{W}$  behaves like a conditional normal random variable, but below behaves like a discrete random variable with two mass points.
Note that, with the previous structure, for values of $\tilde{w}$ less than $\hat{w}$:
\begin{eqnarray*}
P(\tilde{W}\leq \hat{w}, \tilde{V}\leq \tilde{v}|Z=z)&=& P(\tilde{W}\leq \tilde{w}| \tilde{V}\leq \tilde{v},Z=z)\Phi(\tilde{v}) \\
&=&  P(\tilde{W}\leq \hat{w}| \tilde{V}\leq \tilde{v})\frac{e^z}{1+e^z}\Phi(\tilde{v})\\
&=& \int_{-\infty}^{\tilde{v}}P(\tilde{W}\leq \hat{w}| \tilde{V}=v)\phi(v) dv\frac{e^z}{1+e^z}
\end{eqnarray*}
 
Furthermore, note that: 
\begin{eqnarray*}
 \int_{-\infty}^{\tilde{v}} \Phi(\hat{w}|\tilde{v};\rho)\phi(\tilde{v}) d\tilde{v}&=& \int_{-\infty}^{\tilde{w}}\int_{-\infty}^{\tilde{v}}   \phi(\hat{w}|\tilde{v};\rho)\phi(\tilde{v}) d\tilde{v}d\tilde{w}= \int_{-\infty}^{\tilde{w}}\int_{-\infty}^{\tilde{v}}   \phi(\hat{w},\tilde{v};\rho)d\tilde{v}d\tilde{w}\\
 &=& \Phi(\tilde{w},\tilde{v};\rho)
\end{eqnarray*}
With this, we can then say that: 
\begin{eqnarray*}
  F_{\tilde{W}, \tilde{V}|Z}(\tilde{w}, \tilde{v}|z)= \begin{cases}
\Phi(\hat{w},\tilde{v};\rho)\frac{e^z}{1+e^z} \text{ if } \tilde{w} = \hat{w}, -\infty<\tilde{v}  \\
\Phi(-\beta_0-\beta_1z,\tilde{v};\rho)\text{ if } \hat{w}<\tilde{w} \leq -\beta_0-\beta_1z,  -\infty<\tilde{v} \\
 \Phi(\tilde{w},\tilde{v};\rho)\text{ if } \tilde{w}>-\beta_0+\beta_1z,  -\infty<\tilde{v}\\
 0, \text{ otherwise}
 \end{cases}      
\end{eqnarray*}
This joint c.d.f., which integrates to $1$ when $\tilde{w}$ and $\tilde{v}$ goes to $+\infty$ is associated with the following mixed discrete-continuous "density": 
\begin{eqnarray*}
  f_{\tilde{W}, \tilde{V}|Z}(\tilde{w}, \tilde{v}|z)= \begin{cases}
\Phi(\hat{w},\tilde{v};\rho)\frac{e^z}{1+e^z} \text{ if } \tilde{w}= \hat{w}, \forall \tilde{v}  \\
\Phi(-\beta_0-\beta_1z,\tilde{v};\rho)-\Phi(\hat{w},\tilde{v};\rho)\frac{e^z}{1+e^z}\text{ if }\tilde{w}= -\beta_0-\beta_1z,  \forall \tilde{v} \\
 \phi(\tilde{w},\tilde{v};\rho)\text{ if } \tilde{w}>-\beta_0+\beta_1z, \forall \tilde{v}
 \end{cases}
\end{eqnarray*}
Assume that equations \ref{IVTEQ1}-\ref{IVTEQ3} hold. 
Then, define $\bar{Y}_z=\beta_0+\beta_1 z+\tilde{W}, \tilde{D}_z=\gamma_0+\gamma_1 z+\tilde{V}$

\begin{eqnarray}
P(\bar{Y}_z< 0, D_z\leq d_0) &=&  P(\tilde{W}\leq -\beta_0-\beta_1z, \tilde{V}\leq d_0-\gamma_0-\gamma_1z )\\
&=& \Phi_{W,V}( -\beta_0-\beta_1 z,d_0-\gamma_0-\gamma_1 z;\rho)\\
&=& P(Y=0,  D \leq d_0| Z=z)
\end{eqnarray}
The last equality is due to the testable equalities holding. A similar display holds for any $c_0,c_1>0, d_1,d_0$.   Thus a similar logic holds for the other equalities. So this is an example of a model that satisfies all the equalities, and is observationally equivalent to the IV-Tobit, even though the underlying latent error distributions does not satisfy the conditions for the IV-Tobit. Hence, we cannot falsify the assumed IV-Tobit model.
\end{document}